\documentclass{aa}

\usepackage{graphicx}
\usepackage{txfonts}
\usepackage{amsmath}    % Advanced maths commands
\usepackage{amssymb}    % Extra maths symbols
\usepackage{color}
\usepackage{ulem}  %only needed in preparation; remove upon submission

\begin{document}

\title{Forming disc galaxies in major mergers II. The central mass
concentration problem and a comparison of GADGET3 with GIZMO}
\titlerunning{Forming disc galaxies in major mergers.}

\author{S.A.~Rodionov, E.~Athanassoula \and N.~Peschken}

\institute{
Aix Marseille Univ, CNRS, LAM, Laboratoire d'Astrophysique de
Marseille, Marseille, France\\ 
\email{sergey.rodionov@lam.fr}
}
 
\date{Received XX XX, XXX; accepted XX XX, XXXX}
\abstract
{
In a series of papers, we study the major merger of two disk galaxies in order to establish
whether or not such a merger can produce a disc galaxy.
}
{
{Our aim here is to describe in detail the technical aspects of our numerical
experiments. } 
}
{
We discuss the initial conditions of our major merger, which
consist of two protogalaxies on a collision orbit. We show that such merger
simulations can produce a non-realistic central mass concentration,
and we propose simple, parametric, active Galactic nuclei (AGN)-like feedback as a solution to
this problem. Our AGN-like feedback algorithm is very simple: at each
time-step we take all particles whose
local volume density is above a given threshold value and
increase their temperature to a preset value . We also compare
  the GADGET3 and GIZMO codes, by 
  applying both of them to the same initial conditions.
}
{
We show that the evolution of isolated protogalaxies resembles the
evolution of disk galaxies, thus arguing that our protogalaxies are
well suited for our merger simulations. We demonstrate that the problem 
with the unphysical central mass concentration in our merger
simulations is further aggravated when we increase the resolution. 
We show that our AGN-like feedback removes this non-physical
central mass concentration, and thus
allows the formation of realistic bars. Note that our
AGN-like feedback mainly affects the central region of
a model, without 
significantly modifying the rest of the galaxy. 
We demonstrate that, in the context of our kind of simulation, GADGET3 gives
results which are very similar to those obtained with the PSPH
(density independent SPH) flavor of GIZMO. 
Moreover, in the examples we tried, the differences between the results
of the two flavors of GIZMO 
-- namely PSPH, and MFM (mesh-less algorithm) -- are similar to and, in some
comparisons, larger than 
the differences between the results of GADGET3 and PSPH.
}
{}


\maketitle

%%%%%%%%%%%%%%%%% BODY OF PAPER %%%%%%%%%%%%%%%%%%

\section{Introduction}

A number of observational works have 
argued that a large fraction of present-day spiral galaxies have experienced a major merger
event during the last 8 Gyr \citep[e.g.,][]{Hammer.FEZLC.05, Hammer.FPYARD.09,
Hammer.FYAPRP.09, Puech_2012}. This has motivated a number of numerical
simulations modeling the merging and its remnant. 
However, most dynamical simulations of major mergers between two
spiral galaxies  made so far have initial conditions where each
galaxy has the properties of nearby disk galaxies, with, in the best cases, an
enhanced gas fraction to mimic disk galaxies at intermediate redshifts  
\citep{Barnes.02, Springel.Hernquist.05, 
Cox.JPS.06,Hopkins.CYH.09, %Governato.BBMWJSPCWQ.09, 
Lotz.JCP.10b,Wang.HAPYF.12, Hopkins.CHNHM.13, Borlaff.EMRPQTPGZGB.14,
Querejeta.EMTBRPZG.15,
Hopkins.CYH.09,Governato.BBMWJSPCWQ.09,
Hopkins.CHNHM.13, Borlaff.EMRPQTPGZGB.14,
Querejeta.EMTBRPZG.15}. 
Merger remnants from such simulations have, in general, a B/T (bulge to total
stellar mass) ratio that is too high to adequately represent spiral galaxies. 
This can be easily explained assuming that a disk can be formed only
from the gas that survived the merger \citep{Hopkins.CYH.09}. In order
to have a small B/T ratio, it is necessary that a large fraction of gas
survives the merger event. For example,  in the most favorable case of 100\% gaseous progenitors, for $B/T \sim 0.3,$ at least
70\% of the gas must survive the
merger.
But this is relatively difficult to accomplish if all the gas is initially
in the disk. The surface density of the gas in such gas-rich progenitors
is high, so if we simply adopt the Kennicutt-Schmidt law, we find that the star-formation efficiency will be great (i.e., the gas consumption time-scale will be
small). Moreover, star formation will be strongly enhanced during the merger
event \citep{Larson_Tinsley_1978}. As a consequence, a
significant fraction of the gas will not  
survive the merging, and will turn into stars before its end, which will
lead to high B/T fraction. 

In A16, we proposed a
more realistic setup for the dynamical study of a major merger event
between two spiral galaxies, introducing gas in the form of a gaseous halo, existent in
each of the protogalaxies.  
We thus collide not a couple of fully evolved local spiral
galaxies, but a couple of protogalaxies, consisting of only dark
matter and gas halos. Each protogalaxy, after $1-2$
Gyr of evolution in isolation, resembles an intermediate-redshift
disk galaxy, while 
after $8-10$~Gyr of evolution, it resembles a present-day spiral
galaxy (see A16
and Appendix~\ref{sec:isolated}).
In the current article, the second
of a series, we discuss in detail the technical aspects of our numerical
experiments. In section~\ref{sec:sim}, we describe the initial
conditions for our simulations, while in
Appendix~\ref{sec:isolated} we discuss the evolution of individual
protogalaxies in isolation. In
sect. \ref{sec:agn}, we demonstrate why we need AGN-like feedback in our
models, describe the adopted feedback, and test it. In
sect.~\ref{sec:GIZMO}, we compare results of our simulations
calculated using either GADGET3 or GIZMO codes. 
We conclude in section~\ref{sec:conclusion}.
\section{Simulations}
\label{sec:sim}

\subsection{Initial conditions}
\label{subsec:ic}

\subsubsection{Individual protogalaxies}
\label{subsubsec:individual_ic}

In all our simulations, individual protogalaxies are initially
spherical and composed of a dark matter (DM) halo and a
gaseous halo, of masses $M_{DM}$ and $M_{gas}$,
respectively. Stars do not exist in the initial conditions, but form
during the evolution. 
To within a multiplicative constant, the type of functional form for
the density of both the DM and the gaseous halo is

\begin{equation}
\rho(r) =  \frac{C~~sech(r/|r_t|)}{x^{\gamma_i}~~(x^\eta +
1)^{(\gamma_{o} - \gamma_{i})/\eta}},
\label{eq:halodens}
\end{equation}

\noindent
where $r$ is the spherical radius, $x=r/r_{s}$ , and $r_{s}$ and $r_{t}$ are
characteristic radii of the halo that can be considered as
measures of the scale length and the tapering radius,
respectively. The constants $\gamma_{i}$,
$\gamma_{o}$ , and $\eta$ characterize the radial density
profile and $C$ is a global multiplicative constant to set the
total mass to the desired value. For the DM, in this article, we use
$\gamma_i=1$, $\gamma_o=3,$ and $\eta=1$, which corresponds, up to the
truncation, to Navarro–Frenk–White (NFW) models \citep{NFW1996, NFW1997}. For the
gas, we have $\gamma_i=0$, $\gamma_o=2, $ and $\eta=2$ (Beta
model for $\beta=2/3$, e.g.,~\citet{Miller_Bregman_2015} and
references therein). All parameters for each model discussed here can
be found in Appendix~\ref{s:parameters}. 

 The halo component was built as a spherical isotropic system in equilibrium with the
total potential including gas, using the distribution function technique
\citep[\S 4.3]{BT_2008}.
For this, we used the program mkhalo written by W. Dehnen \citep{McMillan_Dehnen_2007}, which is part of the
NEMO tool box \citep{Teuben_1995}.  We also have the possibility to
add a spin
to the halo. The fraction of particles with a positive sense of rotation
is given by the parameter $f_{pos}$. In cases with no net rotation $f_{pos}=0.5$,
while if all particles rotate in the positive (negative) sense this
parameter is equal to 1 (-1).

The gaseous component was constructed as follows. For each gas particle, all
 velocities except the tangential
velocity $v_{\phi}$ were set to zero. The tangential velocity was
set to the value of $\bar v_{H,\phi}$ ,  the mean tangential velocity in
the halo at the location of the gas particle. This value can
be calculated directly from the distribution function of the halo. 
First, at a given spherical radius $r$, we calculate $\bar v(r)$,
the mean of the module of the velocity vector.

\begin{equation}
\bar v(r) = \frac {\int_0^{v_{max}} v p_v(v) \mathrm{d}v} {\int_0^{v_{max}} p_v(v) \mathrm{d}v},
\end{equation}
 
\noindent
where $v_{max}=\sqrt{2 \Psi}$ is the maximal velocity at a given radius
$r$, $\Psi$ is the negative of the total gravitational potential at a
given radius,  
$p_v(v) \sim v^2 {\rm DF}(\Psi(r) - v^2 / 2)$ is the probability density of
$v$ at a given radius and ${\rm DF}$ is the distribution function \citep[\S 4.3]{BT_2008}.  
It is easy to prove that in 3D space, if we start from an isotropic system and add
rotation by changing the fraction of particles with positive and negative
sense of rotation (by introducing $f_{pos}$), the mean of the tangential
velocity $\bar v_{\phi} = (f_{pos} - 0.5) \bar v$.

It would have been possible to add spin to the system
in a different way. For example $f_{pos}$ in the halo does not have to be
constant and could be set as a function $f_{pos}(R,z)$, where $R$ is the
cylindrical radius. Also, rotation in the gas could be set as completely
independent of the rotation in the halo.

The internal energy of each gaseous particle
is calculated so that it is in hydrostatic equilibrium in the halo + gas
potential, taking into account the spin, if it exists. In the case
without rotation, from the condition
of hydrostatic equilibrium and assuming that the pressure is zero at
infinity, we have: 
\begin{equation}
\label{e:hydrostatic}
P(r') = \int_{r'}^\infty G \rho(r) M(r) / r^2 \mathrm{d}r,
\end{equation}
where $M(r)$ is the total mass inside a sphere with radius $r$,
and $\rho(r)$ and $P(r)$ are the density and pressure, respectively, of the gas at a given
radius. From the pressure,
assuming the condition of ideal gas, we can calculate the specific internal
energy $u=P/(\rho~(\gamma - 1))$, where 
$\gamma$ is the adiabatic index (in all our simulations, we set
$\gamma=5/3$, which corresponds to mono-atomic gas). 

Whenever present, the spin is taken into account
using  the following simple approximation. 
We derive the thermal energy in the polar regions $u_{p}$ from
equation~(\ref{e:hydrostatic}), and calculate the thermal energy in the
equatorial region $u_{e}$  from the pressure:
\begin{equation}
P_{e}(r') = \int_{r'}^\infty  \rho(r) (G M(r) / r^2 - v^2_{\phi}/r) \mathrm{d}r
.\end{equation}
We interpolate the thermal energy as, $u= k u_{p} + (1 - k) u_{e}$,
where $k=|\theta|/ (\pi/2)$, and $\theta$ is the angle
between a given direction and the equatorial plane.
However, we should note that in all models considered here, the
correction for the spin is almost
negligible and it would be enough to use equation~(\ref{e:hydrostatic})
without any corrections.

 Models constructed in such a way are in equilibrium if the gas is
adiabatic (i.e., without cooling and feedback). During the
runs, however, the gas will cool radiatively and, getting out
of equilibrium, fall inwards.
In Appendix~\ref{sec:isolated} we discuss the evolution of such
protogalaxies in detail.

\subsubsection{Collisions}
\label{subsubsec:collisions_ic}

In all cases considered here, we have mergers between two equal-mass
galaxies. However, cases with unequal galaxies
can be easily created in our framework, and they will be discussed
in following articles.

Each of the two protogalaxies is initially created with $Z$ as rotation
axis. After that, they can be arbitrary orientated, which, in general,
gives us two Euler angles per protogalaxy. Let 
the plane of the collisional orbit be the $X-Y$ plane.
We describe this orbit by an initial separation in phase space
$x_0$, $y_0$, $v_{x,0}$, $v_{y,0}$. In some cases, we choose the initial
separation from a two-body problem (assuming that all the mass of the
protogalaxies is concentrated in two points), setting the desired orbit eccentricity,
initial separation, and distance in pericenter. See all the parameters of the
models discussed here in Appendix~\ref{s:parameters}.

\subsection{Code}
\label{subsec:code}

We use a version of GADGET3 including gas and its physics
\citep{gadget2_2005}. The stars and the dark
matter are followed by N-body particles and gravity is
calculated with a tree code. The code uses an improved SPH
method \citep{Springel_Hernquist_2002} and sub-grid
physics \citep{Springel_Hernquist_2003}. 
We use the same parameters for the sub-grid physics as
\citet{Springel_Hernquist_2003}, which were calibrated with Kennicutt's
law \citep{Kennicutt_1998}).  

Except for test simulations, we use a softening length of 25 pc for the gas
and stars and of 50 pc for the halo, a 0.005 relative accuracy of the
force, and the GADGET
opening criterion 1, that is, a relative criterion that tries to limit the
absolute truncation error of the multiple expansion for every
particle-cell interaction (see GADGET manual,
http://www.mpa-garching.mpg.de/gadget/users-guide.pdf).
We also use the GADGET system of units; that is, the
unit of length is 1 kpc, the unit of mass is 10$^{10}$ $M_\odot$ , and
the unit of velocity is 1 km/sec. We continued all simulations up to 10
Gyr.

\section{Central mass concentration in merger models}
\label{sec:agn}

\subsection{Models without AGN-like feedback}
\label{ss:woutagn}

\subsubsection{The central mass concentration problem}
\label{sss:cmc_problem}

As shown in A16 and the following papers, our simulations
can make galaxies whose properties are generally in good agreement
with those of real galaxies. Such comparisons include face-on and
edge-on morphologies, radial density
profiles, finding type II profiles with inner and outer disc
scale lengths, and break radii in good agreement with those observed, etc.. We
also find good agreement with observations for vertical density profiles and thick disk
properties, as well as for
a number of kinematic properties. In particular, their rotation curves
are flat, as observed by \citet{Bosma_1981}.

\begin{figure*}
\includegraphics[width=\linewidth]{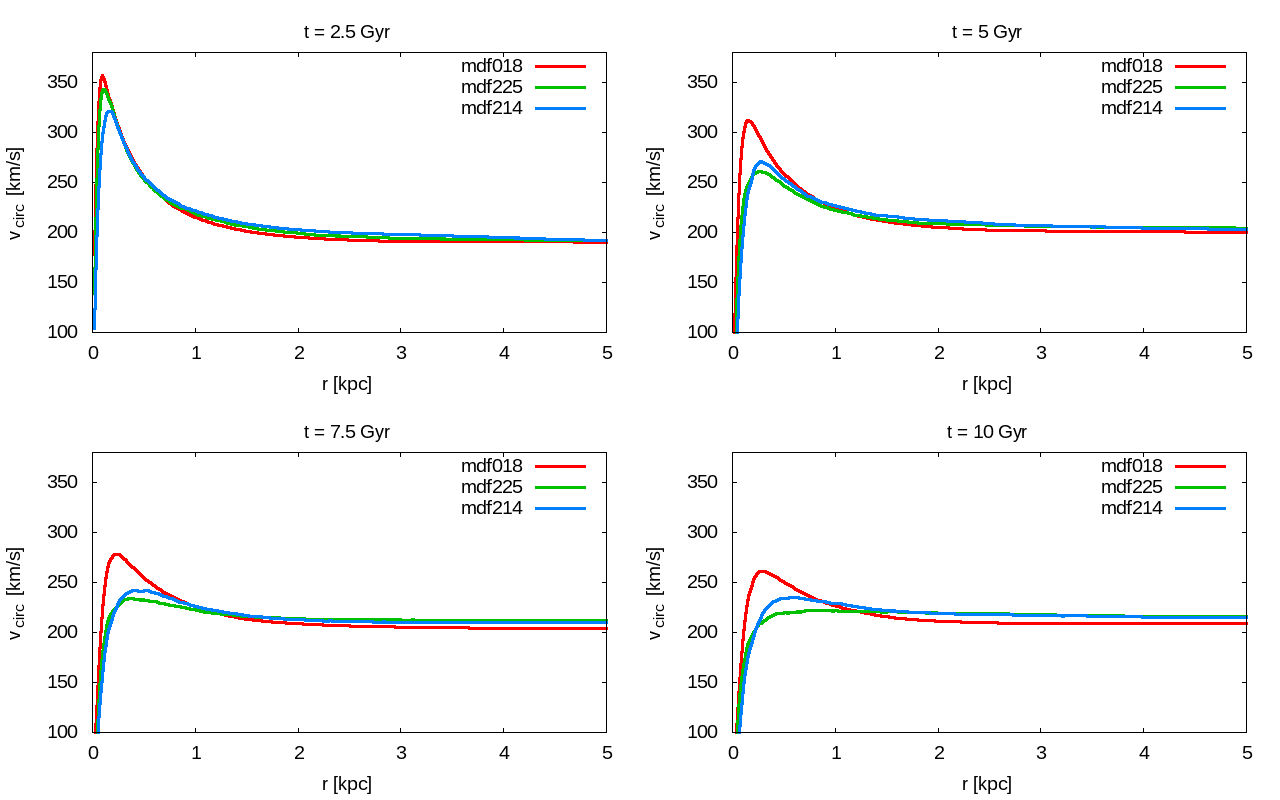}
\caption{Central part of the circular velocity profiles for models mdf018
(high resolution model), mdf225 (low resolution model),
and mdf214 (low resolution models with two times bigger
softening) at times 2.5, 5, 7.5, and 10 Gyr. This is calculated from 
the total mass distribution, assuming spherical symmetry.}
\label{f:vcirc_018_225_214}
\end{figure*}

However, our models without AGN feedback have one notable problem,  
concerning the very central part of the disk. During the collision, the
models form an extremely dense, and, as we discuss below, non-realistic central mass concentration (CMC).
This central mass concentration can be better seen on the circular
velocity profile $v_c(R)$\footnote{ 
In this paper we calculate
the circular velocity from the total mass distribution, assuming spherical
symmetry: $v_c(R)=\sqrt{G~M(R)/R}$, where $M(R)$ is the total mass
inside a sphere
of radius $R$. }.

%We have chosen to do so because here we are
%interested in the differences of the mass distributions in the central region,
%and $v_c(R)$ calculated in such a way is a direct function of mass.

Let us consider a typical example of a model without AGN: mdf018. All
parameters of this model can be found in Appendix~\ref{s:parameters}. 
This model is the non-AGN analog of mdf732, which was thoroughly analyzed
in A16. 
In Fig.~\ref{f:vcirc_018_225_214} (red line), we show the central part
of the $v_c(R)$ profiles at different times
for mdf018. At $t=2.5$~Gyr, that is, approximately 1 Gyr after the merger,
the central bump on $v_c(R)$ is very high, and reaches almost $360$ km/s (see
fig.~\ref{f:vcirc_018_225_214}). 

 The strong central peak on $v_c(R)$ is unrealistic by itself.
The circular velocity curves of the models can be compared with observed HI
or H$\alpha$ rotation curves. Central bumps on rotation
curves may appear in real galaxies,
but are considerably smaller \citep[e.g.,][]{Sofue_1999}. Another problem with the
strong central peaks on $v_c(R)$ concerns bars. Our simulations show
that such a strong central peak stops, or
delays beyond $10$~Gyr, the formation of a bar (see also A16), while approximately two thirds of
disk galaxies have bars \citep{Buta_2015}. 
Bumps of similar size did  also appear in
cosmological simulations until relatively recently, but have been
lately strongly diminished
by increasing the feedback \citep[see][for a
review]{Brooks_2016}. We will follow a similar route (see
sect.~\ref{ss_agn_how}), 
but before doing so, we would like to continue the discussion of
CMCs in a model without an AGN.

\subsubsection{Demise of the CMC in low-resolution models.}

In Fig.~\ref{f:vcirc_018_225_214}, we can see that during the evolution, the central
bump on the $v_c(R)$ profile in mdf018 becomes less and less prominent
with time, and at $t=10$~Gyr it has only a maximum of $260$~km/s. 
One can suggest that this is due to two-body relaxation
\citep{BT_2008}. In order to demonstrate this, we consider two low-resolution test models, each with ten times less particles than
mdf018 (see Appendix~\ref{s:parameters}). One model (mdf225) has the same gravitational softening as
mdf018: 50 pc for the halo, and 25 pc for gas and stars. The other
model (mdf214) has double this softening: 100 pc for the halo, and 50 pc for gas and stars.  
All parameters, except the number of particles and the softening
lengths, are identical to those of mdf018. Low-resolution models with
the same gravitational softening (mdf225) at $t=2.5$~Gyr have a central
peak on $v_c(R)$ with almost the same maximum as in the high-resolution model mdf018 (see
Fig.~\ref{f:vcirc_018_225_214}, green line). Nevertheless, during the evolution,
the central bump decreases much faster than in mdf018, and by $t=10$~Gyr,
the low resolution model mdf225 has no central bump on $v_c(R)$ at all. This fits
well with the fact that the relaxation time is smaller for a smaller number of
particles. Low-resolution models with a softening twice as big (mdf214), at
$t=2.5$~Gyr, have a central peak significantly less pronounced on
$v_c(R)$, which is expected due to the decreased spacial resolution.
During the evolution, however, the central bump on $v_c(R)$ decreases
slower than in low-resolution simulations with a smaller softening (mdf225), which is
also expected, because the relaxation time is bigger for a bigger softening
\citep{Athanassoula_2001, RS_2005}. This analysis argues that the lowering
of the central peak on $v_c(R)$ in the high-resolution model mdf018 is
due, at least partly, to the two-body relaxation.  

 As a result, we can expect that higher-resolution simulations, namely
with a greater number of particles
and smaller softening, will make the problem with CMC even worse,
because a
smaller softening will make the peak on $v_c(R)$ higher, and a bigger number
of particles will prevent its attenuation with time. On the other hand,
smaller resolution spuriously diminishes the problem (see Fig.~\ref{f:vcirc_018_225_214}, for two low-resolution models, mdf214 and mdf225). This
means that, in sufficiently low-resolution models, one will not have
problems with a non-realistic CMC. Of course, decreasing the resolution
cannot be considered as a solution to the high central concentration
problem.

\subsection{AGN-like feedback. How we will proceed. }
\label{ss_agn_how}

One could expect that adding AGN feedback would solve, or at least
alleviate, the problem with
the central peak in the circular velocity profile. However, an AGN is not trivial to model.
There are inherent difficulties in understanding both the accretion into the
central black hole (hereafter BH) \citep{Hopkins_Quataert_2010} and the subsequent energetic
feedback \citep{King_2003,Wurster_Thacker_2013}. Moreover, we obviously
do not have enough resolution to model these two processes directly. 

Nevertheless, there are several AGN feedback algorithms (see, for
example, a comparative study of \citet{Wurster_Thacker_2013}, and papers
by \citet{Dubois_2010, Blecha_2013, Volonteri_2015, Gabor_2016}). Usually, such
algorithms include the explicit calculation of the accretion onto the black hole (thus
the evolution of the mass of the black hole), its movement (advection),
and the merging of black holes in cases of collisional simulations.
But here  we propose a much simpler and less ambitious approach. We do
not aim to calculate the evolution of the super-massive black holes; we only
want to remove non-physical central mass concentrations in our models by
adding physically motivated feedback.
 One way to proceed in such a situation is to introduce
some kind of subgrid-physics model and ``calibrate'' it with
observations. 

Our intent was to include, in our model, a simple, parametric, AGN-like feedback with the
following properties:
\begin{itemize}
\item 
Being simple.
\item
Being able to solve the problem with the central peak in the circular velocity profile.
\item
Allowing the formation of a bar.
\item
Being physically plausible, thus injecting a reasonable amount of energy in the
central region of the galaxy. 
\item
Influencing mainly the central region of the model.
\end{itemize}

 The last point requires some explanation. It is commonly believed
that AGN feedback could be quite strong and, presumably, be able to significantly
quench star formation in the entire galaxy
\citep{Springel_et_al_2005, Di_Matteo_2005, Bundy_2008}, thereby influencing more than
just the 
central region of the galaxy. However, as we mentioned before, 
there are a lot of fundamental
difficulties in proper modeling of AGN feedback. So, if we include an AGN-feedback that modifies the entire galaxy  in
our model, we will need to
calibrate it with some observations, which is not trivial. Because of
this, we have decided that it would be reasonable, as a first step, to
include an AGN-like feedback that mainly influences the
central region of the model, and barely affects the outer parts. 

\subsection{Description of our AGN-like feedback}
\label{ss:agn_description}

 Our AGN-like feedback is based on two parameters: a volume density threshold
$\rho_{AGN}$, and a temperature $T_{AGN}$. More specifically, at every
time step, we give internal energy to the gas particles whose local
volume density
is larger than the threshold $\rho_{AGN}$, by increasing their
temperature to $T_{AGN}$. The density threshold is chosen so as to
ensure that the particles are located in the center-most region.

 In our algorithm, we do not have a single particle representing
the black hole, so we do not directly follow the mass of the
BH. We can, however, 
calculate the BH mass from the feedback energy by, in some sense, 
inverting the formalism described in \citet{Springel_et_al_2005}.
We take the same value as \citet{Springel_et_al_2005} for the
radiative efficiency of the BH, $\epsilon_r=0.1$, and
the fraction of radiative luminosity $\epsilon_f=0.05,$ which can couple
thermally to the surrounding gas.
Consequently: 
\begin{equation}
M_{\rm BH}(t)=E_{feed}(t) / (c^2 \epsilon_f \epsilon_r) + M_{\rm BH}(0),
\end{equation}
where $M_{\rm BH}(t)$ is the BH mass at a given time t, and $E_{feed}(t)$
is the cumulated feedback energy up to time t. We stress that
in the basic version 
of our feedback algorithm (without Eddington limit), we do not need to know
$M_{\rm BH}$ in the algorithm itself.

The Eddington limit can be added as an option. It requires three new free
parameters: $\epsilon_r$, $\epsilon_f$, and $M_{\rm BH}(0)$. 
Knowing $M_{\rm BH}(t),$ we can calculate the Eddington luminosity as:
\begin{equation}
L_{\rm Edd} = \frac{4  \pi G  M_{\rm BH}(t)  m_p c }{\sigma_{\rm T}}.
\end{equation}
From this, we can calculate the energy available for the feedback for the current
time-step $dt$ as $E_{\rm Edd} = L_{\rm Edd} \cdot dt \cdot
\epsilon_f$. We then calculate
$E_{req}$, the energy required to heat all particles with $\rho > \rho_{AGN}$
up to a temperature of $T_{AGN}$. If $E_{req} > E_{\rm
  Edd}$ , we heat particles
with a probability $p = E_{\rm Edd} / E_{req}$. This way,
we statistically obey the Eddington limit. For further 
discussion of the Eddington limit see Sect. \ref{ss:eddington}.

\subsection{Models with and without AGN-like feedback}

Here we compare five models; all of them have identical
parameters, except for the AGN-like feedback parameters; 

\begin{itemize}
\item
mdf018 : Model without AGN-like feedback
\item
mdf789 : $T_{AGN}=5 \times 10^6~K$, $\rho_{AGN}=1~M_\odot / pc^3$ 
\item
mdf732 : $T_{AGN}=1 \times 10^7~K$, $\rho_{AGN}=1~M_\odot / pc^3$ 
\item
mdf788 : $T_{AGN}=2 \times 10^7~K$, $\rho_{AGN}=1~M_\odot / pc^3$ 
\item
mdf791 : $T_{AGN}=1 \times 10^7~K$, $\rho_{AGN}=2~M_\odot / pc^3$ 
\end{itemize}
The remaining parameters are given in Appendix~\ref{s:parameters}.

\begin{figure*}
\includegraphics[width=\linewidth]{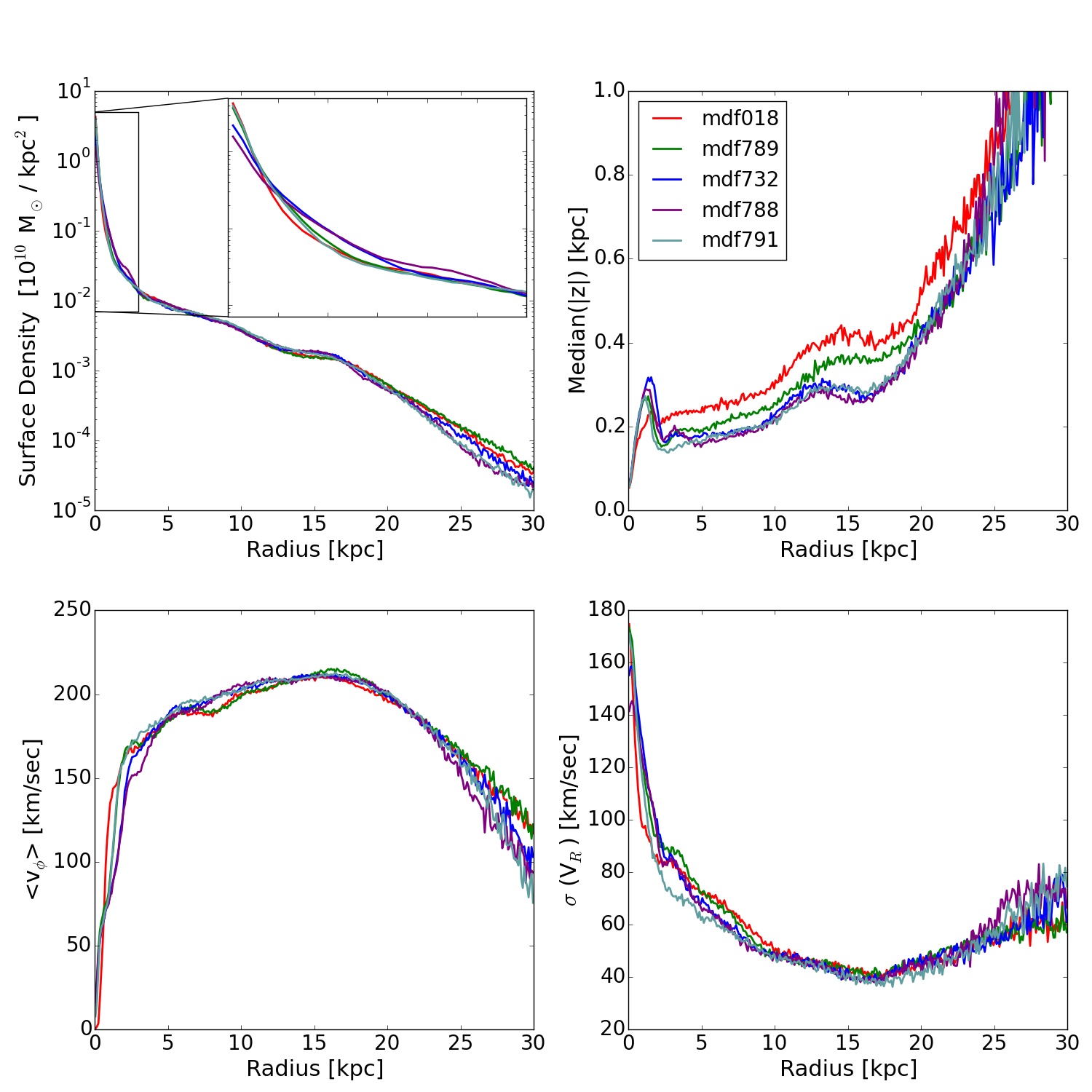}
\caption{Comparison of various radial profiles for the stellar
component of five models with different AGN feedback parameters:
mdf018 (no AGN), mdf789 ($T_{AGN}=5 \times 10^6~K$,
$\rho_{AGN}=1~M_\odot / pc^3$), mdf732 ($T_{AGN}=1 \times 10^7~K$,
$\rho_{AGN}=1~M_\odot / pc^3$), mdf788 ($T_{AGN}=2 \times 10^7~K$, $\rho_{AGN}=1~M_\odot / pc^3$), 
and mdf791 ($T_{AGN}=1 \times 10^7~K$, $\rho_{AGN}=2~M_\odot / pc^3$),
all at t=10 Gyr.
From left to right and top to bottom: surface density,
median of the absolute value of z, which is a good approximation for
thickness \citep{Sotnikova_Rodionov_2006}, mean value of the azimuthal
velocity and radial velocity dispersion. The inlay in the upper left
panel shows the surface density in the innermost region (within 3 kpc).}
\label{f:combine1_y32_family}
\end{figure*}

\begin{figure*}
\includegraphics[width=\linewidth]{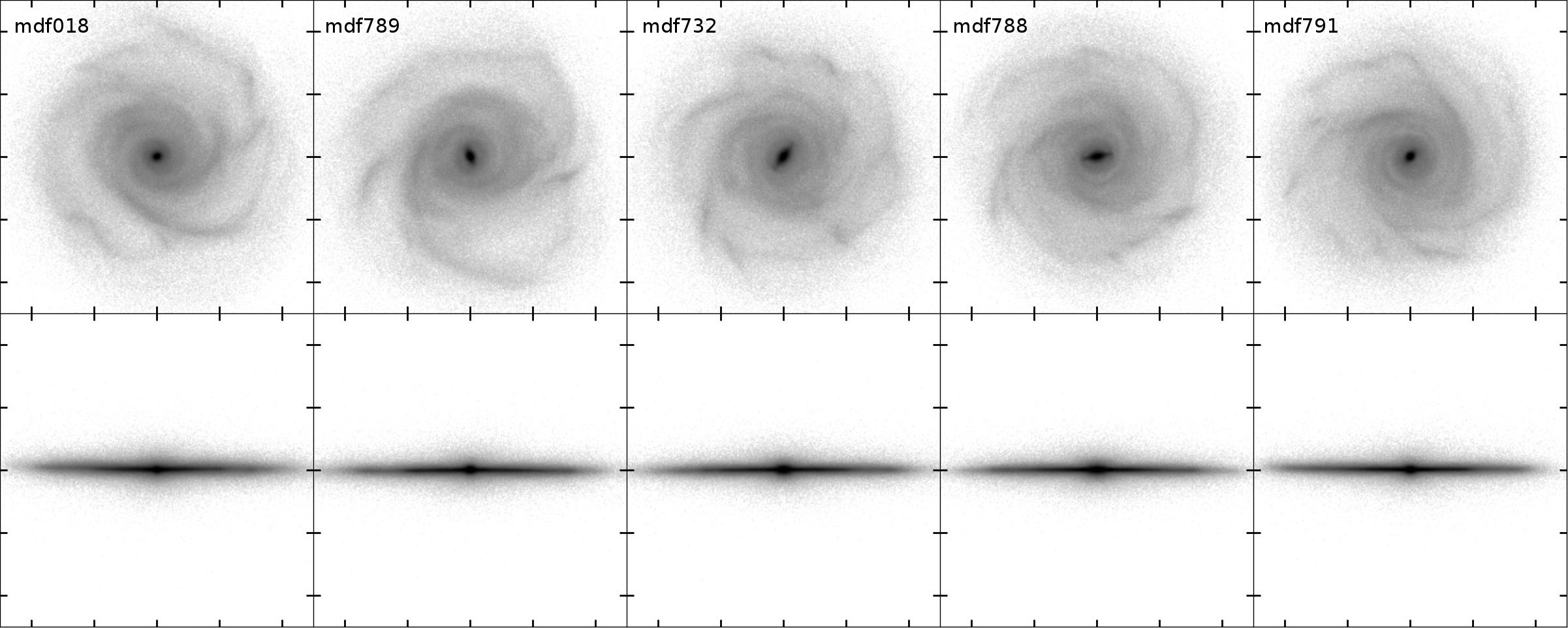}
\caption{Face-on (upper row) and edge-on (lower row) views of the stars component for 
five models with different AGN feedback parameters:
  mdf018 (no AGN), mdf789 ($T_{AGN}=5 \times 10^6~K$,
  $\rho_{AGN}=1~M_\odot / pc^3$), mdf732 ($T_{AGN}=1 \times 10^7~K$,
  $\rho_{AGN}=1~M_\odot / pc^3$), mdf788 ($T_{AGN}=2 \times 10^7~K$, $\rho_{AGN}=1~M_\odot / pc^3$), 
and mdf791 ($T_{AGN}=1 \times 10^7~K$, $\rho_{AGN}=2~M_\odot / pc^3$),
from left to right, respectively, and all at t=10 Gyr. The size of each square box corresponds to 50 kpc.}
\label{f:view_y32_family}
\end{figure*}

\begin{figure*}
\includegraphics[width=\linewidth]{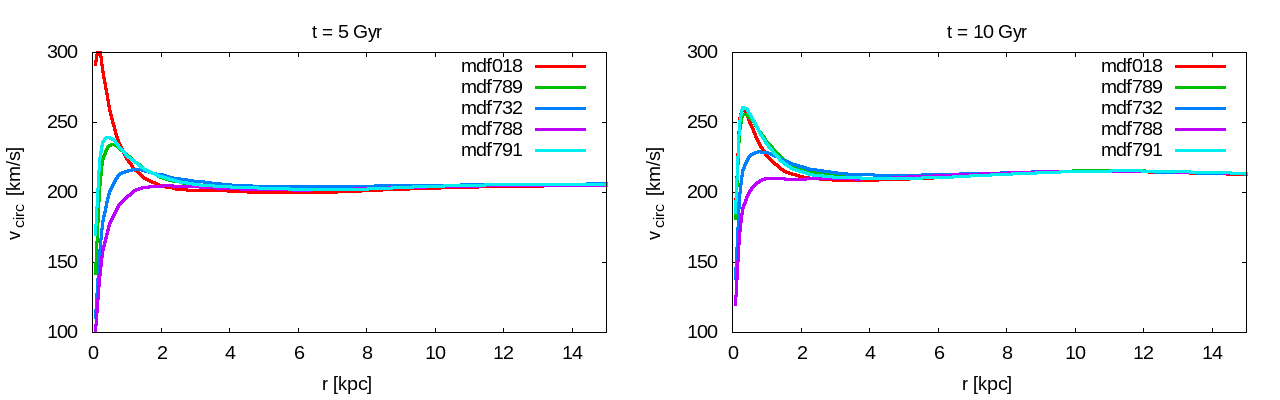}
\caption{Circular velocity profiles for five models with
different AGN feedback parameters:
  mdf018 (no AGN), mdf789 ($T_{AGN}=5 \times 10^6~K$,
  $\rho_{AGN}=1~M_\odot / pc^3$), mdf732 ($T_{AGN}=1 \times 10^7~K$,
  $\rho_{AGN}=1~M_\odot / pc^3$), mdf788 ($T_{AGN}=2 \times 10^7~K$,
  $\rho_{AGN}=1~M_\odot / pc^3$) and 
mdf791 ($T_{AGN}=1 \times 10^7~K$, $\rho_{AGN}=2~M_\odot / pc^3$), at
5 (left) and 10 (right panel) Gyr.}
\label{f:vcirc_y32_family}
\end{figure*}

At t=$10$~Gyr, all these models are very similar. 
Their radial surface density profiles
are virtually identical, except for the innermost region where models
with no or  weak feedback (mdf018, mdf789, and mdf791) have a steeper
cusp than the others (fig. \ref{f:combine1_y32_family}).  It is the
extra density in their cusps that is responsible for the steep maximum in
the circular velocity curve, already discussed in Sect.~\ref{sss:cmc_problem}.
Kinematic profiles including the mean azimuthal velocity and the
radial velocity dispersion profiles are also very similar
(Fig.~\ref{f:combine1_y32_family}). However, comparing face-on views, we see 
differences in morphology (fig.~\ref{f:view_y32_family}). 
Despite the fact that azimuthally averaged profiles are very
similar, in the outer part of the models, we can see relatively strong differences in
the shape of the spiral structure. These differences are partly due to
the transient nature of the spirals, but we can not exclude that they
are also partly due to differences in the central
part. Anyway, the spiral structure has often been referred to as ``the
frosting on the cake''. Below, we will discuss the central part of the models.

 First, let us analyze the model without AGN (mdf018). As already
discussed in Sect.~\ref{sss:cmc_problem}, during the collision, an extremely dense and presumably
non-realistic CMC is created. This 
can be better seen on the circular velocity profile ($v_c(R)$) and
is very prominent at $t=5$~Gyr (see Fig.~\ref{f:vcirc_y32_family} for the
model without AGN, mdf018). At
$t=10$~Gyr, this peak is less prominent, which could partly be due to
the numerical two-body relaxation, but a part of it, albeit
small, could perhaps also reflect true two-body relaxation, as in 
globular clusters.
Because of this strong central mass concentration, we do not have a bar
in this model (model mdf018, without
AGN, see fig.~\ref{f:view_y32_family}). 

Let us now consider a sequence of models with fixed $\rho_{AGN}$, and increasing
$T_{AGN}$: mdf789 ($T_{AGN}=5 \times 10^6~K$), mdf732 ($T_{AGN}=1
\times 10^7~K$), and mdf788 ($T_{AGN}=2 \times 10^7~K$). This can be considered as a
sequence with increasing feedback strength. The weak AGN in mdf789
($T_{AGN}=5 \times 10^6~K$) manages to  only partly remove the very dense CMC (see
Fig.~\ref{f:vcirc_y32_family}, $t=5$~Gyr), which, however, is still relatively
prominent, and at $t=10$~Gyr, the $v_{c}$ profile is
similar to that of mdf018 (Fig.~\ref{f:vcirc_y32_family}, $t=10$~Gyr). At this
time, we do have a bar, but it is quite small, with a semi-major axis of approximately 1.5 kpc at 10 Gyr (see
Fig.~\ref{f:view_y32_family} for mdf789). In mdf732, we increase $T_{AGN}$
to $10^7$~K, and in this model, the central bump on the $v_{c}$ profile is much
less prominent, which allows the formation of a bigger bar 
with a semi-major axis of approximately 3 kpc (see
fig.~\ref{f:vcirc_y32_family} and fig~\ref{f:view_y32_family} for
mdf732). 
A further increase of $T_{AGN}$ to $2 \times 10^7$ in mdf788 fully
removes the central bump on $v_{c}$, but the bar size in this model is
similar to mdf732 ($T_{AGN}=1 \times 10^7~K$). We must note that one can expect the relation
between the bar and the AGN
to be more complicated than a simple linear correlation,
and that a strong AGN, by shuffling material in
the central region, could, in some cases, delay or even prevent the formation of
a bar (the analysis of this subject is beyond the scope of this
article). We note that mdf732 is one of our reference models and was
thoroughly analyzed in A16.

Our last model to compare is mdf791 ($T_{AGN}=1 \times 10^7~K$, $\rho_{AGN}=2~M_\odot / pc^3$),
which is similar to mdf732 ($T_{AGN}=1 \times 10^7~K$, $\rho_{AGN}=1~M_\odot / pc^3$), but
has a $\rho_{AGN}$ twice as high as the others. We can see that
by increasing $\rho_{AGN}$ , we decrease the efficiency of the AGN and make this
model very similar to mdf789 ($T_{AGN}=5 \times 10^6~K$,
$\rho_{AGN}=1~M_\odot / pc^3$, see Figs.~\ref{f:view_y32_family} and~\ref{f:vcirc_y32_family}). This can be easily understood because an
increase in $\rho_{AGN}$ means that fewer particles will have their
thermal energy increased and thereby the total feedback will be smaller.

The vertical thickness is a further noticeable difference between the models with and 
without AGN (see fig.~\ref{f:combine1_y32_family}). In the very
central region (less than 2 kpc), the difference in the thickness is due to
the presence of a bar, and particularly to the boxy/peanut bulge associated to it. But
the differences outside the bar region are more difficult to explain.  
The model with a very
weak AGN (mdf789), beyond the bar region, is approximately in between the model without AGN and the rest
of the models. Presumably it is somehow connected to the very dense central
mass concentration but the exact mechanism which causes such
a difference is not clear for us yet.

In general, we can say that our simple AGN-like feedback is exactly
what we wanted it to be. It introduces the desired considerable changes in the
central part of the model, as aimed for, and only relatively small
ones in the rest of the galaxy. More specifically, it solves the
problem with the central peak of the $v_c$ profile, and allows the 
formation of the bar.

\subsection{Models with and without Eddington limit}
\label{ss:eddington}

The main reason for including the Eddington limit in an AGN feedback
model is to make sure that the amount of feedback energy is reasonable.
There is, however, both observational and theoretical evidence for the
possibility of supercritical accretion \citep{King_2003, Sadowski_2014}.
So even from a general point of view, including Eddington limits may not
be obligatory in order to be physically reasonable.

Here we will show that in our simple, parametric, AGN-like feedback
model, the Eddington limit can be almost fully compensated by varying $T_{AGN}$. 

Let us compare three models. These models have identical
parameters, except for $T_{AGN}$ and the Eddington limit:
\begin{itemize}
\item 
mdf732, with Eddington limit, $T_{AGN}=1 \times 10^7~K.$
\item
mdf751, with Eddington limit, $T_{AGN}=1.5 \times 10^7~K.$
\item
mdf726, without Eddington limit, $T_{AGN}=1 \times 10^7~K.$
\end{itemize}
The remaining parameters for these models are given in Appendix~\ref{s:parameters}.

\begin{figure*}
\includegraphics[width=\linewidth]{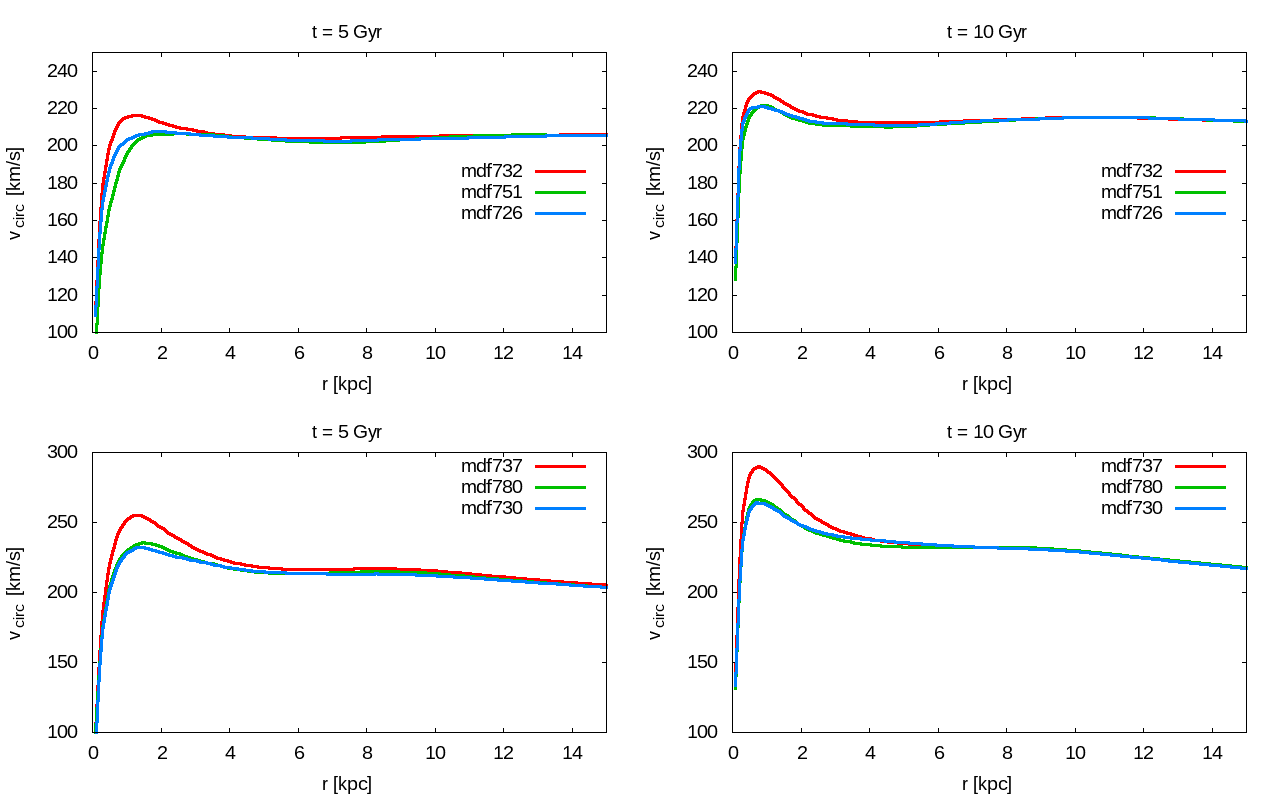}
\caption{Comparison of circular velocity profiles (at 5 and 10 Gyr) for models with
and without Eddington limits. The upper row shows two models with Eddington limit (mdf732 $T_{AGN}=1 \times 10^7~K$ and
mdf751 $T_{AGN}=1.5 \times 10^7~K$) and one model without Eddigton limit 
(mdf726 $T_{AGN}=1 \times 10^7~K$). The lower panels also display
three models: two with Eddigtion limit (mdf737 $T_{AGN}=2 \times 10^7~K$,
and mdf780 $T_{AGN}=2.5 \times 10^7~K$) and one model without (mdf730
$T_{AGN}=2 \times 10^7~K$).}
\label{f:vcirc_El_nonEL}
\end{figure*}

All three models are very similar, although there is a small difference
in the mass distribution in the central part (see upper panels of 
fig.~\ref{f:vcirc_El_nonEL}). If we compare mdf732 and mdf726, which
differ only in the presence, or absence of the Eddington limit, we can see
that in a model without Eddington limit (mdf732), the AGN is somewhat
less efficient and the central bump on the
circular velocity profile is more prominent, which is expected
because of the Eddington limit. However, when we make a simulation with
Eddington limit, but increase the value of
$T_{AGN}$ by 50\%, we fully compensate for the presence of the
Eddington limit (see model mdf751 on
fig.~\ref{f:vcirc_El_nonEL}).

Let us consider another set of three models:
\begin{itemize}
\item
mdf737, with Eddington limit, $T_{AGN}=2 \times 10^7~K.$
\item
mdf780, with Eddington limit, $T_{AGN}=2.5 \times 10^7~K.$
\item
mdf730, without Eddington limit, $T_{AGN}=2 \times 10^7~K.$
\end{itemize}
The remaining parameters for these models are given in Appendix~\ref{s:parameters}.

These models have a more eccentric orbit and a larger initial
separation than the previous ones. This leads to a greater
amount of low-angular-momentum
gas in the central part, immediately after the collision, and consequently requires a more energetic AGN (bigger $T_{AGN}$) in order to remove
the central peak in the circular velocity profile and make the bar formation possible.  
Nevertheless, this does not change the conclusion of this section. If we compare 
models with and without the Eddington limit, but with the same
$T_{AGN}$, then we observe that the former has a larger central bump
(compare mdf737, which has an Eddington limit, to mdf730, which does not,
in the lower panels of Fig.~\ref{f:vcirc_El_nonEL}). It is, however, possible to
compensate for this by simply increasing $T_{AGN}$ (see mdf780 with
Eddington limit and $T_{AGN}=2.5 \times 10^7~K$ vs. mdf730 without
Eddington limit and $T_{AGN}=2 \times 10^7~K$ in lower panels of Fig.~\ref{f:vcirc_El_nonEL}).
Thus, it is possible to compensate for the effect of the Eddington limit on
the circular velocity curve by varying the AGN-like feedback parameters.

\section{Comparison with GIZMO}
\label{sec:GIZMO}
\subsection{Introductory remarks} 

The GIZMO simulation code \citep{Hopkins.13, Hopkins.14, Hopkins.15} is a
fork of GADGET3 (hereafter G3), using the same MPI parallelization, 
domain decomposition, gravity solvers, and so on. However, 
in contrast to G3, GIZMO offers several hydrodynamical
solvers for the user to choose from. In this article, we compare the
three solvers,
which are briefly described below (see \citet[sect. 4.1]{Hopkins.15}
for more information).
\begin{itemize}
\item
TSPH (``Traditional'' SPH): this is the same, entropy conserving,
density-driven formulation of SPH as the one in G3.
\citep{Springel_Hernquist_2002}.
\item
PSPH: this is a density independent version of SPH
\citep{Hopkins.13}. It also includes a better 
treatment of the artificial viscosity term
\citep{Cullen.Dehnen.10}, as described in \citeauthor{Hopkins.13}
 (\citeyear{Hopkins.13}, sect. 3.1).
\item
MFM (Meshless Finite-Mass): Lagrangian formulation of mesh-free
algorithm that conserves particle masses \citep{Hopkins.15}.
\end{itemize}

Phase boundaries, such as, for example, those  
between the hot gas in the halo and the much colder gas in
the disk, or those around cold dense clumps of gas in the
hot halo can be
better handled by PSPH or MFM, than by TSPH or G3 \citep[see][and references therein]{Hopkins.15}. 

Since our aim here is to see how results will change with different
hydro-solvers, we keep the same subgrid physics.

The main difference we found between G3 and PSPH/MFM is in the hot gaseous halo. 
This has small gaseous clumps in the
simulation run with G3, which are either absent,
or much less prominent in the PSPH or MFM runs (see also
\citealt{Torrey.VSSH.12} and \citealt{Hayward.TSHV.14}).
However, in our project, we are mostly interested in the properties of
the remnant. Therefore, in this comparative study, we
will focus on the properties of the final galaxy.

\subsection{Simulations without AGN-like feedback}

We first compared G3 to TSPH and, as expected, we found no
differences. We will thus not discuss TSPH further here.

Let us now compare three simulations of major mergers without AGN. All
three simulations were run with the same parameters, those of mdf018
(see~Appendix~\ref{s:parameters}),
except for the softening, which is 50 pc for both dark matter and
baryons, but with a different 
hydrodynamical solver. We also used the same
subgrid physics in all three cases. Thus, the only difference is the hydro-solvers. These
three runs are mdf624  
(run with G3), mdi101 (run with GIZMO/PSPH), and 
mdh101  (run with GIZMO/MFM). 

%\begin{figure*}
%\includegraphics[width=\linewidth]{figs/GIZMO_nonAGN_AGN_view_xy.png}
%\caption{Comparison of face-on views of the stars component for G3,
%PSPH and MFM models, at t=10 Gyr.
%First row, models with-out AGN: mdf624 (G3), mdg101(PSPH), mdi101(MFM). 
%Second row models with AGN: mdf627 (G3), mdg115 (PSPH), mdi115 (MFM).
%The size of each square box corresponds to 50 kpc.}
%\label{f:GIZMO_view_xy}
%\end{figure*}

\begin{figure*}
\includegraphics[width=\linewidth]{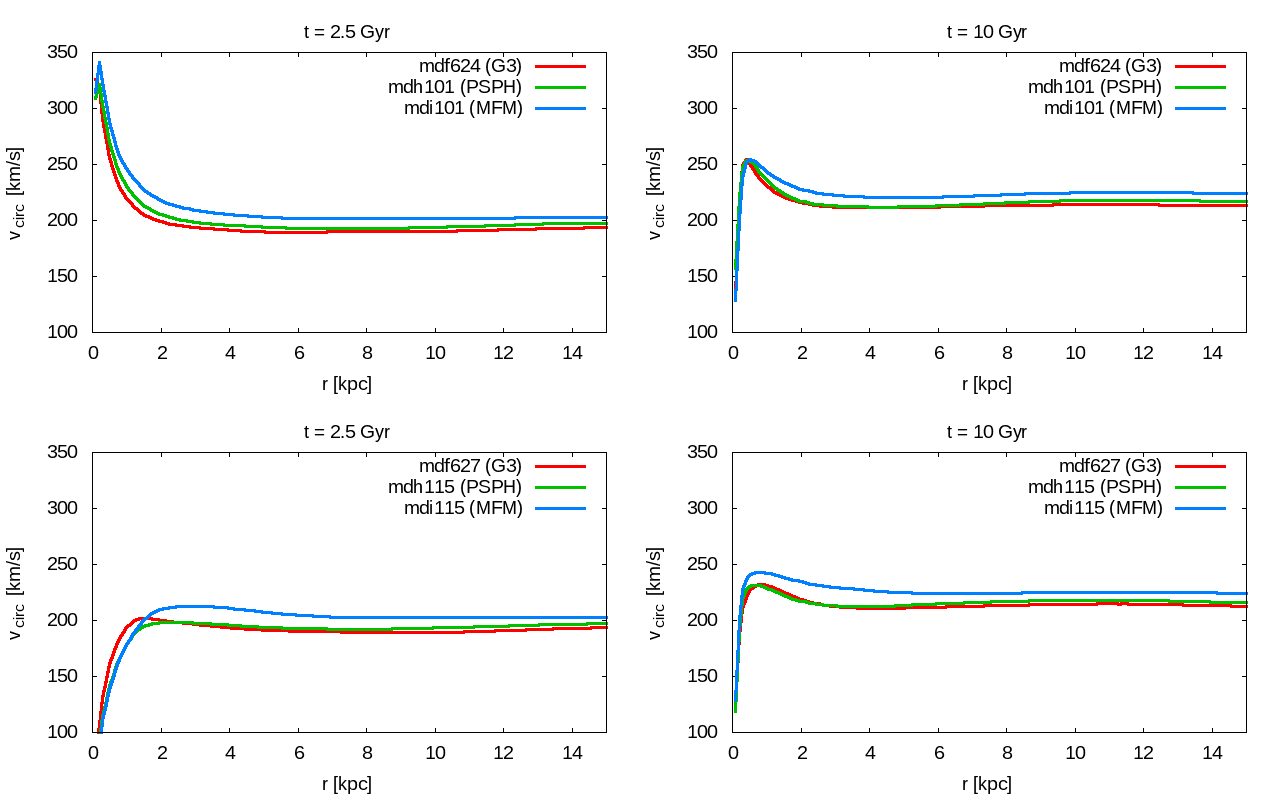}
\caption{ Comparison of circular velocity profiles for G3, PSPH, and
MFM models at $t$=2.5 Gyr (left panels) and $t$=10 Gyr (right panels).
The upper row compares models without AGN %, namely mdf624 (G3),
%mdg101(PSPH) and mdi101(MFM). 
and the lower row compares models with AGN. %: mdf627 (G3), mdg115
%(PSPH), mdi115 (MFM). 
} 
\label{f:GIZMO_vcirc}
\end{figure*}

\begin{figure*}
\includegraphics[width=\linewidth]{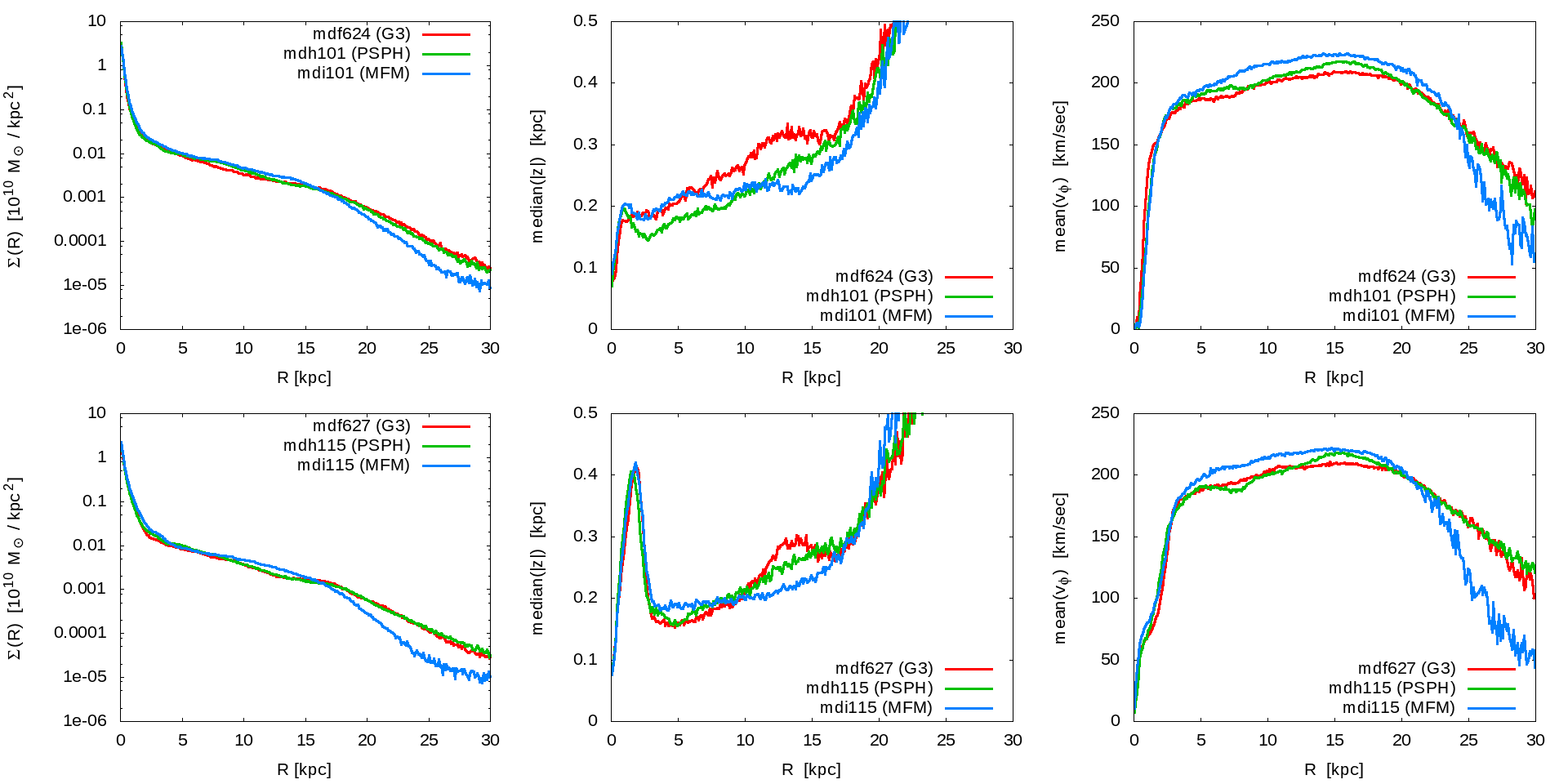}
\caption{
Comparison of various radial profiles for G3, PSPH, and MFM models.
The upper row shows models without AGN %: mdf624 (G3), mdg101(PSPH), mdi101(MFM). 
and the lower row models with AGN. %: mdf627 (G3), mdg115 (PSPH), mdi115 (MFM).
In each row, from left to right: surface density,
median of the absolute value of z, which is a good approximation for
thickness \citep{Sotnikova_Rodionov_2006} and mean value of the azimuthal
velocity}
\label{f:GIZMO_combine2}
\end{figure*}

Comparisons are made in the upper panels of Figs.~\ref{f:GIZMO_vcirc}
(for the circular velocity curves) and \ref{f:GIZMO_combine2} (for
the projected surface density, the vertical thickness, and the mean
tangential velocity of the stars). In the former, we see that at $t$=2.5 Gyr, all 
three runs have a very strong CMC; that of the MFM run being somewhat
stronger than the other two. Nevertheless, in all three cases, the peak is
considerably higher than 300 km/sec, for a velocity of the flat part
of the circular velocity curve of approximately 200 km/sec. By $t$=10 Gyr,
this CMC has considerably decreased, as expected (see
sect.~\ref{sss:cmc_problem}), but
still has a maximum of approximately 250 km/sec, still giving an unrealistic
shape of the circular velocity curve, although much less so than that
at $t$=2.5 Gyr. In general, the G3 and the PSPH runs
give very similar results, in most cases differing by approximately no more than the
plotting accuracy. On the other hand, MFM gives somewhat higher
values, corresponding to a somewhat higher mass. We note that the CMC
found in these runs, which have no AGN-like feedback, is sufficient to
prohibit bar formation, or at least to delay it beyond the 10 Gyrs over
which we make the comparisons. 

In the upper left panel of Fig.~\ref{f:GIZMO_combine2}, we compare the radial
projected surface density profiles, as obtained at $t=10$~Gyr for the three
simulations. It is clear that the inner disc scale length (i.e., the
scale length of the part of the disc which is within the break in the
projected surface density, around $R=17$~kpc) is approximately
the same for all three models. The outer disc scale length
is approximately the same for models G3 and PSPH, while for MFM it is somewhat smaller, leading to a
faster drop in the surface densities in the outer disc region
(i.e., the part of the disc beyond the break). The middle panel
compares the radial profile of a measure of the stellar disk vertical
thickness, namely the  azimuthal averaged median of the absolute value of the $z$
coordinates of all stellar particles \citep{Sotnikova_Rodionov_2006}. The three
methods give similar results, and, although the differences are larger
than what we found for the circular velocities and the projected
surface densities, they are still relatively small, especially taking into
account that dist thickness is very sensitive to numerical effects
\citep{Rodionov_Sotnikova_2013}. We also note that in this
plot, the MFM does not stand out as being further apart from the other
two runs. 

Finally the rightmost panel compares the radial profiles of the
three mean stellar tangential velocity radial profiles. Here, again, the G3 and the PSPH
are very similar and the MFM differs somewhat more, but never by too much.      

\begin{figure*}
\includegraphics[width=\linewidth]{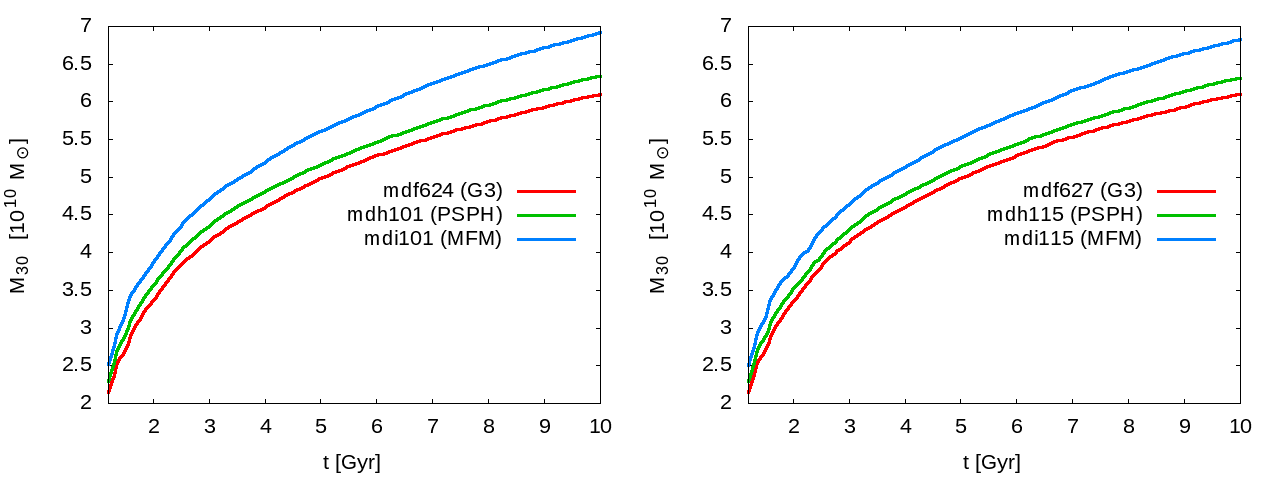}
\caption{ Comparison of the baryonic mass {(gas + stars), within a sphere of
30 kpc radius} as a function of time for G3, PSPH, and
MFM models. The left panel is for simulations without AGN-like
feedback and the right one with it.
} 
\label{f:GIZMO_mbar_30}
\end{figure*}

A further comparison is given in Fig.~\ref{f:GIZMO_mbar_30}, where we
present the evolution of the baryonic mass (gas + stars)
inside a sphere of radius 30 kpc with time. We can see that in the MFM model,
gas accretes on the disk somewhat faster than in the PSPH and G3 models. As
a result, at $t=10$ Gyr, the disk in the MFM models is approximately 12\% more massive.
There is also some difference between the G3 and the PSPH models, but
this is much smaller than the difference between the MFM and the other two
models. Indeed, at $t$=10 Gyr, the difference between G3 and PSPH
amounts to only approximately 4\% of the baryonic mass.  
These differences in accretion rate could explain the corresponding
differences between the three circular velocity curves discussed above.

\subsection{Simulations with AGN-like feedback}

Here, we again first compared G3 with GIZMO in the case of the same
hydro solver (G3 vs. TSPH). We found that, with the same parameters,
our AGN-like feedback is slightly less efficient in TSPH runs than in
G3 runs. It should be noted that even with the TSPH hydro solver, GIZMO is
not identical to G3 \citep[sect.~F1]{Hopkins.15}. 
Nevertheless, this difference can be easily compensated for by tuning the
feedback parameters, and we do so in the following.

Let us now compare three other runs, identical to the three compared
in the previous subsections, but now including the proposed AGN-like feedback. 
\begin{itemize}
\item 
mdf627 : run with GADGET3. $T_{AGN}=1 \times 10^7~K$,
$\rho_{AGN}=1~M_\odot / pc^3$, with Eddigton limit.
\item 
mdh115 : run with GIZMO/PSPH. $T_{AGN}=1 \times 10^7~K$,
$\rho_{AGN}=0.75~M_\odot / pc^3$, without Eddigton limit.
\item
mdi115 : run with GIZMO/MFM. $T_{AGN}=1 \times 10^7~K$,
$\rho_{AGN}=0.75~M_\odot / pc^3$, without Eddigton limit.
\end{itemize}
These simulations were run with  50 pc softening for both dark matter and
baryons. The remaining parameters are those of mdf018 (see~Appendix~\ref{s:parameters})

The results are again compared in Figs.~\ref{f:GIZMO_vcirc}
and \ref{f:GIZMO_combine2}, but now in the lower panels.  
We note that in all three cases this feedback leads to a realistic
circular velocity curve with no strong CMC, contrary to what we found in runs
with no AGN-like feedback, which all had strong CMCs. Furthermore, we witnessed that,
with the demise of the CMC, bars form in all three cases under
consideration. In fact, it is due to these bars that there is a maximum of
the disk thickness in the central region for all three models (lower
middle panels of Fig.~\ref{f:GIZMO_combine2}).
Considering all the comparisons globally, we see that in the case with
AGN-like feedback,
the MFM differs from the other two more than in the cases
without AGN-like feedback. The results of runs with G3 and PSPH are
again very close to each other, similarly to cases without AGN-like
feedback.

\section{Summary and Conclusions}
\label{sec:conclusion}
 In this paper, we discuss, in detail, the technical aspects of our
wet major merger simulations, where we
collide two idealized protogalaxies. In
section~\ref{subsec:ic}, we describe the initial conditions of our simulations.
Each protogalaxy is initially spherical and consists of a dark matter
halo and a gaseous halo. Spin is added to the system. In Appendix A, we demonstrate
that in isolation their evolution resembles the evolution of disk
galaxies; after 1-2 Gyr they tend to be similar to intermediate
redshift disks and after 10 Gyr they resemble present day spirals.

For the simulations, we use the GADGET3 code \citep{gadget2_2005} with
star-formation and feedback modeled by sub-grid physics 
\citep{Springel_Hernquist_2003}. We demonstrate that, without any
additional feedback, our major merger models have a very compact and
massive CMC (see \ref{ss:woutagn}). This leads to an unphysically high
central peak of the circular velocity profile. Moreover, this compact
and massive CMC 
prevents the formation of a bar, while approximately two thirds of real disk galaxies have bars.
We demonstrate that low resolution can, in some cases,
artificially mask this problem. In models with a small number of
particles, a very
compact CMC will be relatively quickly attenuated by two-body relaxation,
and bigger gravitational softening will make CMC less compact from the beginning.
Of course, we cannot consider decreasing the resolution as a
``solution'' to the problem.
We note that higher resolution, namely a larger number of
particles and a smaller softening, will make the CMC problem even more acute.

We solve the problem of the compact CMC by adding simple AGN-like
feedback: at every time step we give internal energy to the gas
particles whose local volume density
is larger than the threshold $\rho_{AGN}$ by increasing their
temperature to $T_{AGN}$. We
demonstrate that this simple AGN-like feedback is able to solve
the problem with the central peak on the circular velocity profile, making this
profile very realistic, and allowing bar formation. We also
show that our AGN-like feedback mainly changes
 the central region of the disk without significantly modifying the rest of the model. 

As an option, in our AGN-like feedback, we
have an Eddington limit (see section~\ref{ss:agn_description}).
However, we demonstrate that the absence or presence of
the Eddington limit in the models we have considered so far can be
compensated, at least as far as the mass and velocity distributions
are concerned, by changing the $T_{AGN}$
parameter. This, together with
observational and theoretical evidence of the possibility of supercritical accretion \citep{King_2003,
Sadowski_2014}, argues that one is not obliged to include an Eddington
limit in the AGN-like feedback described here in order to obtain
models with realistic density and velocity profiles.
They can, however, prove useful if one wants to set
reasonable limits to the energy injected into the central regions.

We have also compared results of our simulations calculating with
three different hydro-solvers: traditional SPH (G3 or TSPH in the GIZMO
code), density
independent SPH (PSPH, GIZMO code), and Lagrangian formulation of mesh-free
algorithm, which conserves particle masses (MFM, GIZMO code).
We found that for the specific results we are interested in,
namely properties of major merger remnants, G3, and PSPH give, in all cases
we tried, very similar results, and one can use either of the two codes.
Moreover, the differences between MFM and PSPH are similar, and, in some
comparisons, even bigger than the differences between PSPH and G3.

\section*{Acknowledgements}

We thank P. Hopkins and V. Springel for providing us with
the versions of GIZMO and GADGET, used here, and for useful
discussions. We also thank an anonymous referee for comments that
helped improve the presentation of this paper. 
We acknowledge financial support from  
CNES (Centre National d’Etudes Spatiales, France) 
and from the EU Programme FP7/2007-2013/, under REA grant
PITN-GA-2011-289313. We also acknowledge
HPC resources from GENCI/TGCC/CINES (Grants x2013047098 and x2014047098) 
and from Mesocentre of Aix-Marseille-Universite (program DIFOMER).

\bibliographystyle{aa}
\bibliography{a2015_2}

\begin{thebibliography}{62}
\expandafter\ifx\csname natexlab\endcsname\relax\def\natexlab#1{#1}\fi

\bibitem[{{Athanassoula} {et~al.}(2016){Athanassoula}, {Rodionov}, {Peschken},
  \& {Lambert}}]{A16}
{Athanassoula}, E., {Rodionov}, S.~A., {Peschken}, N., \& {Lambert}, J.~C.
  2016, \apj, 821, 90

\bibitem[{{Athanassoula} {et~al.}(2001){Athanassoula}, {Vozikis}, \&
  {Lambert}}]{Athanassoula_2001}
{Athanassoula}, E., {Vozikis}, C.~L., \& {Lambert}, J.~C. 2001, \aap, 376, 1135

\bibitem[{{Barnes}(2002)}]{Barnes.02}
{Barnes}, J.~E. 2002, \mnras, 333, 481

\bibitem[{{Binney} \& {Tremaine}(2008)}]{BT_2008}
{Binney}, J. \& {Tremaine}, S. 2008, {Galactic Dynamics: Second Edition}
  (Princeton University Press)

\bibitem[{{Blecha} {et~al.}(2013){Blecha}, {Loeb}, \& {Narayan}}]{Blecha_2013}
{Blecha}, L., {Loeb}, A., \& {Narayan}, R. 2013, \mnras, 429, 2594

\bibitem[{{Borlaff} {et~al.}(2014){Borlaff}, {Eliche-Moral},
  {Rodr{\'{\i}}guez-P{\'e}rez}, {Querejeta}, {Tapia}, {P{\'e}rez-Gonz{\'a}lez},
  {Zamorano}, {Gallego}, \& {Beckman}}]{Borlaff.EMRPQTPGZGB.14}
{Borlaff}, A., {Eliche-Moral}, M.~C., {Rodr{\'{\i}}guez-P{\'e}rez}, C.,
  {et~al.} 2014, \aap, 570, A103

\bibitem[{{Bosma}(1981)}]{Bosma_1981}
{Bosma}, A. 1981, \aj, 86, 1825

\bibitem[{{Bouwens} {et~al.}(2004){Bouwens}, {Illingworth}, {Blakeslee},
  {Broadhurst}, \& {Franx}}]{Bouwens.IBBF.04}
{Bouwens}, R.~J., {Illingworth}, G.~D., {Blakeslee}, J.~P., {Broadhurst},
  T.~J., \& {Franx}, M. 2004, \apjl, 611, L1

\bibitem[{{Brooks} \& {Christensen}(2016)}]{Brooks_2016}
{Brooks}, A. \& {Christensen}, C. 2016, Galactic Bulges, 418, 317

\bibitem[{{Bundy} {et~al.}(2008){Bundy}, {Georgakakis}, {Nandra}, {Ellis},
  {Conselice}, {Laird}, {Coil}, {Cooper}, {Faber}, {Newman}, {Pierce},
  {Primack}, \& {Yan}}]{Bundy_2008}
{Bundy}, K., {Georgakakis}, A., {Nandra}, K., {et~al.} 2008, \apj, 681, 931

\bibitem[{{Buta} {et~al.}(2015){Buta}, {Sheth}, {Athanassoula}, {Bosma},
  {Knapen}, {Laurikainen}, {Salo}, {Elmegreen}, {Ho}, {Zaritsky}, {Courtois},
  {Hinz}, {Mu{\~n}oz-Mateos}, {Kim}, {Regan}, {Gadotti}, {Gil de Paz}, {Laine},
  {Men{\'e}ndez-Delmestre}, {Comer{\'o}n}, {Erroz Ferrer}, {Seibert},
  {Mizusawa}, {Holwerda}, \& {Madore}}]{Buta_2015}
{Buta}, R.~J., {Sheth}, K., {Athanassoula}, E., {et~al.} 2015, \apjs, 217, 32

\bibitem[{{Cox} {et~al.}(2006){Cox}, {Jonsson}, {Primack}, \&
  {Somerville}}]{Cox.JPS.06}
{Cox}, T.~J., {Jonsson}, P., {Primack}, J.~R., \& {Somerville}, R.~S. 2006,
  \mnras, 373, 1013

\bibitem[{{Cullen} \& {Dehnen}(2010)}]{Cullen.Dehnen.10}
{Cullen}, L. \& {Dehnen}, W. 2010, \mnras, 408, 669

\bibitem[{{Daddi} {et~al.}(2010){Daddi}, {Bournaud}, {Walter}, {Dannerbauer},
  {Carilli}, {Dickinson}, {Elbaz}, {Morrison}, {Riechers}, {Onodera}, {Salmi},
  {Krips}, \& {Stern}}]{Daddi.MCMWQ.10}
{Daddi}, E., {Bournaud}, F., {Walter}, F., {et~al.} 2010, \apj, 713, 686

\bibitem[{{Dahlen} {et~al.}(2007){Dahlen}, {Mobasher}, {Dickinson}, {Ferguson},
  {Giavalisco}, {Kretchmer}, \& {Ravindranath}}]{Dahlen.MDFGKR.07}
{Dahlen}, T., {Mobasher}, B., {Dickinson}, M., {et~al.} 2007, \apj, 654, 172

\bibitem[{{Di Matteo} {et~al.}(2005){Di Matteo}, {Springel}, \&
  {Hernquist}}]{Di_Matteo_2005}
{Di Matteo}, T., {Springel}, V., \& {Hernquist}, L. 2005, \nat, 433, 604

\bibitem[{{Dubois} {et~al.}(2010){Dubois}, {Devriendt}, {Slyz}, \&
  {Teyssier}}]{Dubois_2010}
{Dubois}, Y., {Devriendt}, J., {Slyz}, A., \& {Teyssier}, R. 2010, \mnras, 409,
  985

\bibitem[{{Elmegreen} {et~al.}(2005){Elmegreen}, {Elmegreen}, {Vollbach},
  {Foster}, \& {Ferguson}}]{Elmegreen.EVFF.05}
{Elmegreen}, B.~G., {Elmegreen}, D.~M., {Vollbach}, D.~R., {Foster}, E.~R., \&
  {Ferguson}, T.~E. 2005, \apj, 634, 101

\bibitem[{{Erb} {et~al.}(2006){Erb}, {Steidel}, {Shapley}, {Pettini}, {Reddy},
  \& {Adelberger}}]{Erb.SSPRA.06}
{Erb}, D.~K., {Steidel}, C.~C., {Shapley}, A.~E., {et~al.} 2006, \apj, 646, 107

\bibitem[{{Ferguson} {et~al.}(2004){Ferguson}, {Dickinson}, {Giavalisco},
  {Kretchmer}, {Ravindranath}, {Idzi}, {Taylor}, {Conselice}, {Fall},
  {Gardner}, {Livio}, {Madau}, {Moustakas}, {Papovich}, {Somerville},
  {Spinrad}, \& {Stern}}]{Ferguson.P.04}
{Ferguson}, H.~C., {Dickinson}, M., {Giavalisco}, M., {et~al.} 2004, \apjl,
  600, L107

\bibitem[{{Gabor} {et~al.}(2016){Gabor}, {Capelo}, {Volonteri}, {Bournaud},
  {Bellovary}, {Governato}, \& {Quinn}}]{Gabor_2016}
{Gabor}, J.~M., {Capelo}, P.~R., {Volonteri}, M., {et~al.} 2016, \aap, 592, A62

\bibitem[{{Genzel} {et~al.}(2015){Genzel}, {Tacconi}, {Lutz}, {Saintonge},
  {Berta}, {Magnelli}, {Combes}, {Garc{\'{\i}}a-Burillo}, {Neri}, {Bolatto},
  {Contini}, {Lilly}, {Boissier}, {Boone}, {Bouch{\'e}}, {Bournaud}, {Burkert},
  {Carollo}, {Colina}, {Cooper}, {Cox}, {Feruglio}, {F{\"o}rster Schreiber},
  {Freundlich}, {Gracia-Carpio}, {Juneau}, {Kovac}, {Lippa}, {Naab}, {Salome},
  {Renzini}, {Sternberg}, {Walter}, {Weiner}, {Weiss}, \&
  {Wuyts}}]{Genzel.p.15}
{Genzel}, R., {Tacconi}, L.~J., {Lutz}, D., {et~al.} 2015, \apj, 800, 20

\bibitem[{{Governato} {et~al.}(2009){Governato}, {Brook}, {Brooks}, {Mayer},
  {Willman}, {Jonsson}, {Stilp}, {Pope}, {Christensen}, {Wadsley}, \&
  {Quinn}}]{Governato.BBMWJSPCWQ.09}
{Governato}, F., {Brook}, C.~B., {Brooks}, A.~M., {et~al.} 2009, \mnras, 398,
  312

\bibitem[{{Hammer} {et~al.}(2005){Hammer}, {Flores}, {Elbaz}, {Zheng}, {Liang},
  \& {Cesarsky}}]{Hammer.FEZLC.05}
{Hammer}, F., {Flores}, H., {Elbaz}, D., {et~al.} 2005, \aap, 430, 115

\bibitem[{{Hammer} {et~al.}(2009{\natexlab{a}}){Hammer}, {Flores}, {Puech},
  {Yang}, {Athanassoula}, {Rodrigues}, \& {Delgado}}]{Hammer.FPYARD.09}
{Hammer}, F., {Flores}, H., {Puech}, M., {et~al.} 2009{\natexlab{a}}, \aap,
  507, 1313

\bibitem[{{Hammer} {et~al.}(2009{\natexlab{b}}){Hammer}, {Flores}, {Yang},
  {Athanassoula}, {Puech}, {Rodrigues}, \& {Peirani}}]{Hammer.FYAPRP.09}
{Hammer}, F., {Flores}, H., {Yang}, Y.~B., {et~al.} 2009{\natexlab{b}}, \aap,
  496, 381

\bibitem[{{Hayward} {et~al.}(2014){Hayward}, {Torrey}, {Springel}, {Hernquist},
  \& {Vogelsberger}}]{Hayward.TSHV.14}
{Hayward}, C.~C., {Torrey}, P., {Springel}, V., {Hernquist}, L., \&
  {Vogelsberger}, M. 2014, \mnras, 442, 1992

\bibitem[{{Hopkins}(2013)}]{Hopkins.13}
{Hopkins}, P.~F. 2013, \mnras, 428, 2840

\bibitem[{{Hopkins}(2014)}]{Hopkins.14}
{Hopkins}, P.~F. 2014, {GIZMO: Multi-method magneto-hydrodynamics+gravity
  code}, Astrophysics Source Code Library

\bibitem[{{Hopkins}(2015)}]{Hopkins.15}
{Hopkins}, P.~F. 2015, \mnras, 450, 53

\bibitem[{{Hopkins} {et~al.}(2013){Hopkins}, {Cox}, {Hernquist}, {Narayanan},
  {Hayward}, \& {Murray}}]{Hopkins.CHNHM.13}
{Hopkins}, P.~F., {Cox}, T.~J., {Hernquist}, L., {et~al.} 2013, \mnras, 430,
  1901

\bibitem[{{Hopkins} {et~al.}(2009){Hopkins}, {Cox}, {Younger}, \&
  {Hernquist}}]{Hopkins.CYH.09}
{Hopkins}, P.~F., {Cox}, T.~J., {Younger}, J.~D., \& {Hernquist}, L. 2009,
  \apj, 691, 1168

\bibitem[{{Hopkins} \& {Quataert}(2010)}]{Hopkins_Quataert_2010}
{Hopkins}, P.~F. \& {Quataert}, E. 2010, \mnras, 407, 1529

\bibitem[{{Kennicutt}(1998)}]{Kennicutt_1998}
{Kennicutt}, Jr., R.~C. 1998, \apj, 498, 541

\bibitem[{{King}(2003)}]{King_2003}
{King}, A. 2003, \apjl, 596, L27

\bibitem[{{Larson} \& {Tinsley}(1978)}]{Larson_Tinsley_1978}
{Larson}, R.~B. \& {Tinsley}, B.~M. 1978, \apj, 219, 46

\bibitem[{{Leroy} {et~al.}(2008){Leroy}, {Walter}, {Brinks}, {Bigiel}, {de
  Blok}, {Madore}, \& {Thornley}}]{Leroy.WBBBMT.08}
{Leroy}, A.~K., {Walter}, F., {Brinks}, E., {et~al.} 2008, \aj, 136, 2782

\bibitem[{{Lotz} {et~al.}(2010){Lotz}, {Jonsson}, {Cox}, \&
  {Primack}}]{Lotz.JCP.10b}
{Lotz}, J.~M., {Jonsson}, P., {Cox}, T.~J., \& {Primack}, J.~R. 2010, \mnras,
  404, 590

\bibitem[{{McMillan} \& {Dehnen}(2007)}]{McMillan_Dehnen_2007}
{McMillan}, P.~J. \& {Dehnen}, W. 2007, \mnras, 378, 541

\bibitem[{{Miller} \& {Bregman}(2015)}]{Miller_Bregman_2015}
{Miller}, M.~J. \& {Bregman}, J.~N. 2015, \apj, 800, 14

\bibitem[{{Navarro} {et~al.}(1996){Navarro}, {Frenk}, \& {White}}]{NFW1996}
{Navarro}, J.~F., {Frenk}, C.~S., \& {White}, S.~D.~M. 1996, \apj, 462, 563

\bibitem[{{Navarro} {et~al.}(1997){Navarro}, {Frenk}, \& {White}}]{NFW1997}
{Navarro}, J.~F., {Frenk}, C.~S., \& {White}, S.~D.~M. 1997, \apj, 490, 493

\bibitem[{{Puech} {et~al.}(2012){Puech}, {Hammer}, {Hopkins}, {Athanassoula},
  {Flores}, {Rodrigues}, {Wang}, \& {Yang}}]{Puech_2012}
{Puech}, M., {Hammer}, F., {Hopkins}, P.~F., {et~al.} 2012, \apj, 753, 128

\bibitem[{{Querejeta} {et~al.}(2015){Querejeta}, {Eliche-Moral}, {Tapia},
  {Borlaff}, {Rodr{\'{\i}}guez-P{\'e}rez}, {Zamorano}, \&
  {Gallego}}]{Querejeta.EMTBRPZG.15}
{Querejeta}, M., {Eliche-Moral}, M.~C., {Tapia}, T., {et~al.} 2015, \aap, 573,
  A78

\bibitem[{{Rodionov} \& {Sotnikova}(2005)}]{RS_2005}
{Rodionov}, S.~A. \& {Sotnikova}, N.~Y. 2005, Astronomy Reports, 49, 470

\bibitem[{{Rodionov} \& {Sotnikova}(2013)}]{Rodionov_Sotnikova_2013}
{Rodionov}, S.~A. \& {Sotnikova}, N.~Y. 2013, \mnras, 434, 2373

\bibitem[{{Rodrigues} {et~al.}(2012){Rodrigues}, {Puech}, {Hammer}, {Rothberg},
  \& {Flores}}]{Rodrigues.PHRF.12}
{Rodrigues}, M., {Puech}, M., {Hammer}, F., {Rothberg}, B., \& {Flores}, H.
  2012, \mnras, 421, 2888

\bibitem[{{S{\c a}dowski} {et~al.}(2014){S{\c a}dowski}, {Narayan}, {McKinney},
  \& {Tchekhovskoy}}]{Sadowski_2014}
{S{\c a}dowski}, A., {Narayan}, R., {McKinney}, J.~C., \& {Tchekhovskoy}, A.
  2014, \mnras, 439, 503

\bibitem[{{Sofue} {et~al.}(1999){Sofue}, {Tutui}, {Honma}, {Tomita},
  {Takamiya}, {Koda}, \& {Takeda}}]{Sofue_1999}
{Sofue}, Y., {Tutui}, Y., {Honma}, M., {et~al.} 1999, \apj, 523, 136

\bibitem[{{Sotnikova} \& {Rodionov}(2006)}]{Sotnikova_Rodionov_2006}
{Sotnikova}, N.~Y. \& {Rodionov}, S.~A. 2006, Astronomy Letters, 32, 649

\bibitem[{{Springel}(2005)}]{gadget2_2005}
{Springel}, V. 2005, \mnras, 364, 1105

\bibitem[{{Springel} {et~al.}(2005){Springel}, {Di Matteo}, \&
  {Hernquist}}]{Springel_et_al_2005}
{Springel}, V., {Di Matteo}, T., \& {Hernquist}, L. 2005, \mnras, 361, 776

\bibitem[{{Springel} \& {Hernquist}(2002)}]{Springel_Hernquist_2002}
{Springel}, V. \& {Hernquist}, L. 2002, \mnras, 333, 649

\bibitem[{{Springel} \& {Hernquist}(2003)}]{Springel_Hernquist_2003}
{Springel}, V. \& {Hernquist}, L. 2003, \mnras, 339, 289

\bibitem[{{Springel} \& {Hernquist}(2005)}]{Springel.Hernquist.05}
{Springel}, V. \& {Hernquist}, L. 2005, \apjl, 622, L9

\bibitem[{{Tacconi} {et~al.}(2010){Tacconi}, {Genzel}, {Neri}, {Cox}, {Cooper},
  {Shapiro}, {Bolatto}, {Bouch{\'e}}, {Bournaud}, {Burkert}, {Combes},
  {Comerford}, {Davis}, {Schreiber}, {Garcia-Burillo}, {Gracia-Carpio}, {Lutz},
  {Naab}, {Omont}, {Shapley}, {Sternberg}, \& {Weiner}}]{Tacconi.P.10}
{Tacconi}, L.~J., {Genzel}, R., {Neri}, R., {et~al.} 2010, \nat, 463, 781

\bibitem[{{Teuben}(1995)}]{Teuben_1995}
{Teuben}, P. 1995, in Astronomical Society of the Pacific Conference Series,
  Vol.~77, Astronomical Data Analysis Software and Systems IV, ed. R.~A.
  {Shaw}, H.~E. {Payne}, \& J.~J.~E. {Hayes}, 398

\bibitem[{{Toomre} \& {Toomre}(1972)}]{Toomre.Toomre.72}
{Toomre}, A. \& {Toomre}, J. 1972, \apj, 178, 623

\bibitem[{{Torrey} {et~al.}(2012){Torrey}, {Vogelsberger}, {Sijacki},
  {Springel}, \& {Hernquist}}]{Torrey.VSSH.12}
{Torrey}, P., {Vogelsberger}, M., {Sijacki}, D., {Springel}, V., \&
  {Hernquist}, L. 2012, \mnras, 427, 2224

\bibitem[{{Volonteri} {et~al.}(2015){Volonteri}, {Capelo}, {Netzer},
  {Bellovary}, {Dotti}, \& {Governato}}]{Volonteri_2015}
{Volonteri}, M., {Capelo}, P.~R., {Netzer}, H., {et~al.} 2015, \mnras, 449,
  1470

\bibitem[{{Wang} {et~al.}(2012){Wang}, {Hammer}, {Athanassoula}, {Puech},
  {Yang}, \& {Flores}}]{Wang.HAPYF.12}
{Wang}, J., {Hammer}, F., {Athanassoula}, E., {et~al.} 2012, \aap, 538, A121

\bibitem[{{Wurster} \& {Thacker}(2013)}]{Wurster_Thacker_2013}
{Wurster}, J. \& {Thacker}, R.~J. 2013, \mnras, 431, 2513

\end{thebibliography}

\appendix

\section{Isolated models}
\label{sec:isolated}

Here we consider the evolution of three isolated protogalaxies with different
$f_{pos}$ parameters (see sect.~\ref{subsubsec:individual_ic}).
\begin{enumerate}
\item idf407 - $f_{pos}=0.6.$
\item idf401 - $f_{pos}=0.7.$
\item idf413 - $f_{pos}=0.8.$
\end{enumerate}
The remaining parameters are identical for these models and they are given in Appendix~\ref{s:parameters}.

The initial conditions for our protogalaxies are, by construction, in equilibrium for
adiabatic gas (see sect.~\ref{subsubsec:individual_ic}). But, in our
simulations, 
the gas is cooling radiatively during the evolution and, getting
out of equilibrium, falls inwards. Because gas has
angular momentum, it forms a gaseous disk in the central parts, which,
due to star formation, develops a stellar disk component.

\begin{figure*}
\includegraphics[width=\linewidth]{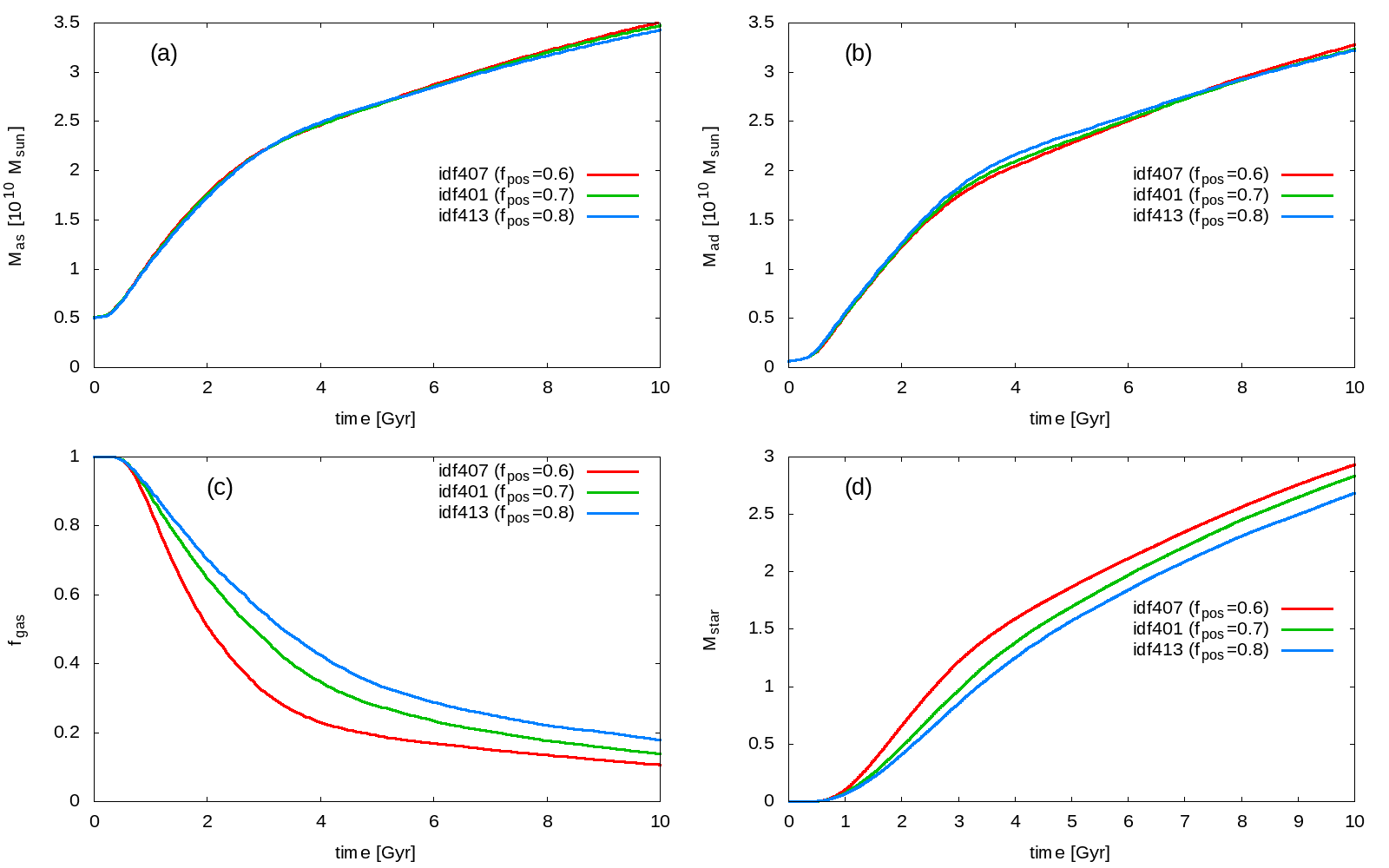}
\caption{Upper row shows cumulative mass of accreted gas as a function
  of time for three isolated models of different initial spin:
(a) shows accretion into 30 kpc sphere and (b) shows accretion into the disk
region with $R<30~{\rm kpc}~~\&~~|z|<2~{\rm kpc}$. Lower panels show:
(c) fraction of gas $f_{gas}=M_{gas}/(M_{gas} + M_{stars})$ 
inside the disk region defined by $R<30~{\rm kpc}~~\&~~|z|<2~{\rm kpc}$ as a function
of time and (d) stellar mass as a function of time}
\label{f:combined1_idf}
\end{figure*}

Before discussing the accreted gas, we need to choose
regions for which we will calculate this accretion. We consider two
regions: a sphere of radius 30 kpc, and a pill-box shaped disk region with
$R<30~{\rm kpc}~~\&~~|z|<2~{\rm kpc}$, where $R$ is the cylindrical
radius. In our three models, only very few stars were formed outside
this disk region (approx. 1.5\%), and even less outside the 30 kpc sphere
(approx. 0.1\%).
In Fig.~\ref{f:combined1_idf}.a we show the cumulative mass of gas
accreted onto the 30 kpc sphere
as a function of time, and in Fig.~\ref{f:combined1_idf}.b we show the 
cumulative mass of gas accreted onto the disk region with
$R<30~{\rm kpc}~~\&~~|z|<2~{\rm kpc}$. The mass of gas accreted onto a
given region was calculated as the mass of baryons (gas + stars)
which, at least once, were inside this region before time $t$.
Actually relatively
few baryons leave the considered regions after they are in, so if we simply use
here the mass of gas+stars inside the given region the results will be very similar.
As can be seen from these figures, the accretion depends very little
on the amount of angular momentum ($f_{pos}$ parameter).
The cumulative accretion onto the 30 kpc sphere for the first $7$~Gyr of
evolution is almost identical for our three 
models, while after this time there is slightly less accretion for models with higher angular momentum
(bigger $f_{pos}$). But when we consider accretion onto the disk itself
(fig.~\ref{f:combined1_idf}.b), the situation is different. At intermediate times, gas
accretes slightly faster onto the disk region for models with higher angular momentum.
However, overall there is little difference, and we 
can conclude that the accretion rate is practically independent on $f_{pos}$.
However, models with smaller angular momentum form stars more
efficiently, as should be expected, because they form more compact
disks. As a result, the gas fraction in the disk
(fig.~\ref{f:combined1_idf}.c), and the stellar mass
(fig.~\ref{f:combined1_idf}.d), depend on $f_{pos}$. Models that at a given time have a smaller
angular momentum (smaller 
$f_{pos}$) have greater stellar masses and smaller fractions
of the gas in the disk region.

\begin{figure*}
\includegraphics[width=\linewidth]{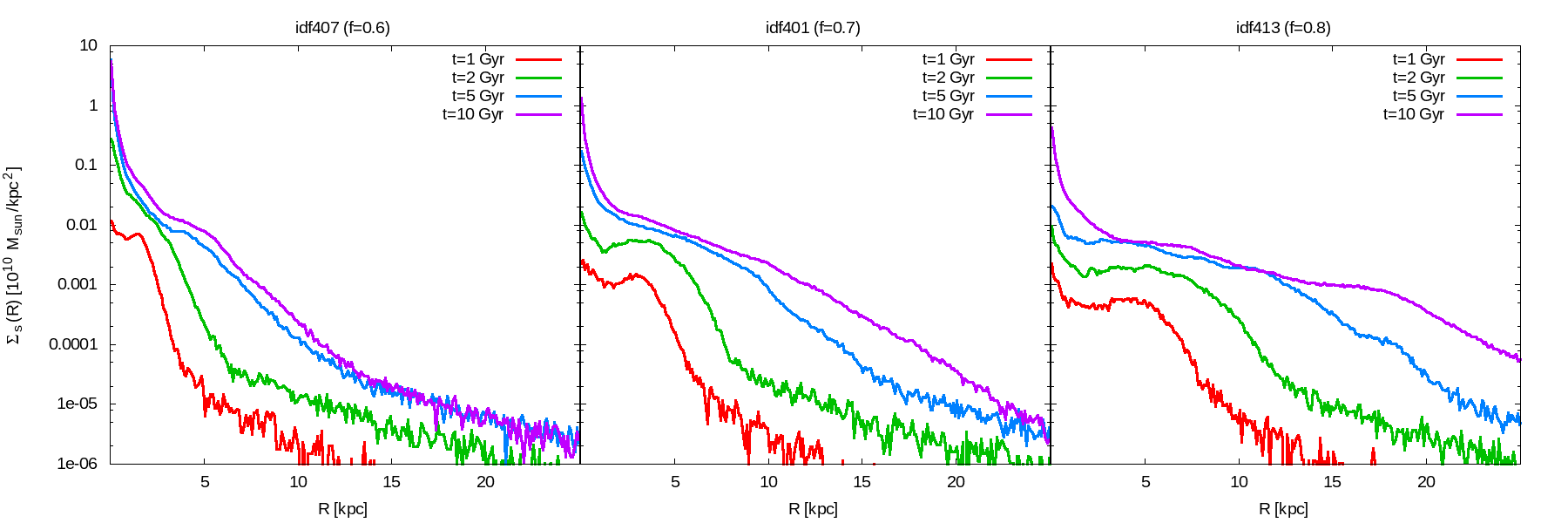}
\caption{Radial stellar surface density profiles for three isolated
models for times 1, 2, 5, and 10 Gyrs.}
\label{f:sd_idf}
\end{figure*}

\begin{figure*}
\includegraphics[width=\linewidth]{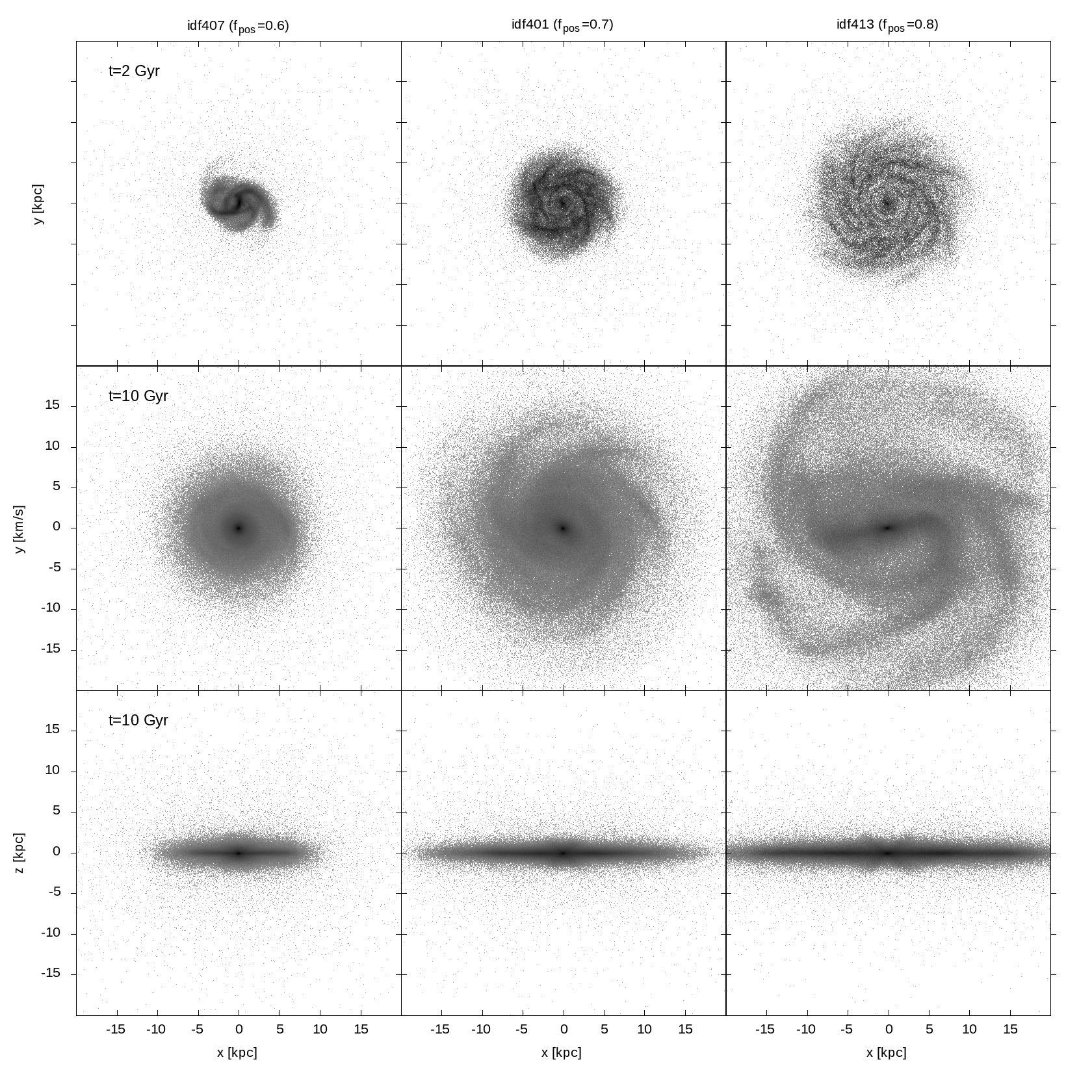}
\caption{Face-on view at t=2 Gyr (upper row), face-on view at t=10 Gyr
(middle row), and edge on view (lower row) at t=10 Gyr for three isolated models:
idf407 ($f_{pos}=0.6$, left column), idf401 ($f_{pos}=0.7$, middle
column) and idf413 ($f_{pos}=0.8$, right column).}
\label{f:view_idf}
\end{figure*}

In Fig.~\ref{f:sd_idf}, we show the surface density profile for three 
models considered at different moments of time. We can see that disks grow
inside-out, and at initial stages, they are significantly smaller than
at $t=10~Gyr$ (see also fig.~\ref{f:view_idf}). At initial stages, our
models are more clumpy and their surface density profiles are, in general,
further away from exponential compared to later stages ($t=10~Gyr$).
This, together with the fact that at early stages the disks in our models
are more gas 
rich (see fig.\ref{f:combined1_idf}.c), makes them similar to
disks in intermediate redshift galaxies, which are considerably
smaller, more perturbed, and have higher gas fractions than disks in
local spiral galaxies \citep[][etc]{Bouwens.IBBF.04, Ferguson.P.04,
Elmegreen.EVFF.05,  Erb.SSPRA.06, Dahlen.MDFGKR.07, Leroy.WBBBMT.08, Daddi.MCMWQ.10,
Tacconi.P.10,  Rodrigues.PHRF.12, Genzel.p.15}.
At $t=10$ Gyr, our models resemble local disk galaxies. They have type~II exponential 
surface density profiles (down-bending) and a fraction of gas of
approximately 10-20\% (see
fig ~\ref{f:combined1_idf}.c). 

\section{Parameters of the models}
\label{s:parameters}

All models considered in this article have the same masses for
the protogalaxies: $M_{DM}=35 \times 10^{10} M_\odot$ and $M_{gas}=5 \times
10^{10} M_\odot$ (see
section~\ref{subsubsec:individual_ic}). The mass of the gas and stellar
particles is $m_g=5 \times 10^4 M_\odot$, and the mass of the DM particles is
$m_{DM}=2 \times 10^5 M_\odot$. Consequently, each
protogalaxy has $10^6$ gas particles and $1.75 \times 10^6$ halo particles.
There is an exception for two low resolution models considered in Sect.~\ref{ss:woutagn}
(mdf214, mdf225), which have protogalaxies with ten times less particles. 
The parameter $f_{pos}$ is fixed to 0.6 for all models with the exception of two
isolated models (for idf401 $f_{pos}=0.7$ and for idf413 $f_{pos}=0.8$
see section~\ref{sec:isolated}), where we wanted to examine the effect
of  $f_{pos}$ on the final density distribution.  

 In all merger models, the two protogalaxies have their rotation axis
parallel to each other and  perpendicular to the orbit of collision 
(see~\ref{subsubsec:collisions_ic}). Here we have two types of
collisional orbits: $o_1$ and $o_2$, whose parameters are given in
Table~\ref{t:orbits}. We note that $o_2$ values were obtained using
the two-body problem 
with the following parameters: eccentricity $\epsilon=0.99$, initial
separation $r_{i}=200$~kpc and distance at pericenter $r_{p}=2$~kpc.

\begin{table}
\caption{Parameters of the collisional orbits.}             
\label{t:orbits}      
\centering                          
\begin{tabular}{c c c c c}       
\hline\hline                 % inserts double horizontal lines
Orbit & $x_0$ [kpc] & $y_0$ [kpc] & $v_{x,0}$ [km/s] & $v_{y,0}$ [km/s] \\
\hline                        
o1  & 100     & 50      & -150 & 0 \\      
o2  & $-198$  & 28.2135 & 125.939  & $-36.6346$ \\
\hline                                  
\end{tabular}
\end{table}

In Table~\ref{t:params}, we present, for all models used here, the
corresponding collisional orbits and the parameters of the AGN-like
feedback. All models with
Eddington limits were calculated with $\epsilon_r=0.1$ and
$\epsilon_f=0.05$, and only the initial mass of the BH $M_{\rm BH}(0)$
is given in Table~\ref{t:params} (see section~\ref{ss:agn_description}).

\begin{table}
\caption{Parameters of the models. }             
\label{t:params}      
\centering                          
\begin{tabular}{c c c c c}       
\hline\hline                 % inserts double horizontal lines
Name & Orbit & $\rho_{AGN}$ [$M_\odot / pc^3$] & $T_{AGN}$ [K] &
$M_{\rm BH}(0)$ [$M_\odot$] \\
\hline                        
mdf018  & o1  & 1  & -                 &   -     \\      
mdf789  & o1  & 1  & $5   \times 10^6$ & $10^5$  \\      
mdf732  & o1  & 1  & $1   \times 10^7$ & $10^5$  \\      
mdf788  & o1  & 1  & $2   \times 10^7$ & $10^5$  \\      
mdf791  & o1  & 2  & $1   \times 10^7$ & $10^5$  \\      
mdf751  & o1  & 1  & $1.5 \times 10^7$ & $10^5$  \\      
mdf726  & o1  & 1  & $1   \times 10^7$ &   -     \\      
mdf737  & o2  & 1  & $2   \times 10^7$ & $10^5$  \\      
mdf780  & o2  & 1  & $2.5 \times 10^7$ & $10^5$  \\      
mdf730  & o2  & 1  & $2   \times 10^7$ &   -     \\      
\hline                                  
\end{tabular}
\end{table}

\end{document}